\documentclass[superscriptaddress,prc,amsfonts,aps,nofootinbib,notitlepage,10pt]{revtex4-1}

\pdfoutput=1
\usepackage{amsmath,amssymb}
\usepackage{graphicx}
\usepackage[utf8]{inputenc}
\usepackage{cancel}
\usepackage[colorlinks=true]{hyperref}
\usepackage{color}
\usepackage{soul}
\usepackage{multirow}
\usepackage{slashed}
\usepackage{enumitem}

\usepackage{multirow}
\usepackage{ulem}

\begin{document}

\title{Systematic shell model study for $N=82$ and $N=126$ isotones and nuclear isomers}

\author{Bharti Bhoy\footnote{bbhoy@ph.iitr.ac.in} and Praveen C. Srivastava\footnote{praveen.srivastava@ph.iitr.ac.in}}
\address{Department of Physics, Indian Institute of Technology Roorkee, Roorkee
247 667, India}

\date{\hfill \today}

\begin{abstract}

 In the present work, we have done a systematic shell model study of $N=82$ and $N=126$ isotones. For the $N=82$ isotones, we have performed calculations using SN100PN interaction, while for $N=126$ isotones, we have used KHPE interaction.
 Similarities between these two isotonic chains have been reported, using the strong resemblance between the high-$j$ orbitals.
 Apart from the nuclear spectroscopic properties, we have also explained different isomeric states in these two regions.
 In the $N$ =82 region, we have mainly discussed the properties of the $6^+$ and $17/2^+$ isomers, while in the $N=126$ region for $8^+$, $11^-$, $21/2^-$ and $29/2^+$ isomers. 
We have reported $B(E2)$, $B(E3)$, $g$-factor, and quadrupole moments of the isomeric states for comparison in these two isotonic chains. 
\end{abstract}

\pacs{21.60.Cs, 21.10.−k, 23.20.-g, 27.60.+j, 27.80.+w}
\maketitle

\section{Introduction}
\label{intro}

The spectroscopy of $^{132}$Sn and $^{208}$Pb and its neighbors in many ways are similar to each other. 
Several nuclear structure studies have been done theoretically and experimentally by comparing the spectroscopy. Some similarities have been pointed out in \cite{Coraggio1,Coraggio2,Coraggio3,Coraggio4,Leander}, showing universal patterns of nuclear structure in both regions. Using the correspondence between the high-$j$ orbitals located above the $Z$ = 50 and $Z$ = 82 shell gaps, Astier et al. \cite{astier1} assigned the isomeric state of $^{217}$Pa as the fully aligned state of the $\pi h_{9/2}^2 f_{7/2}$ configuration. 
They used this analogy to assign the isomeric state of $^{217}$Pa with compared  to the  $J^\pi$= $17/2^+$ isomeric state measured in $^{139}$La.

The single-particle energy gaps in the two cases are comparably large, and the orbitals above and below the shell gaps are similarly arranged \cite{blomqvist}. Every single particle state in the $^{132}$Sn region has its counterpart in the vicinity of $^{208}$Pb. 
Several groups \cite{Coraggio1,Coraggio2,Coraggio3,Coraggio4,Leander,blomqvist,erokhina} have worked to develop a general theoretical description of nuclear structural properties 
 near and far from the $\beta$-stability line using the shell model. They have emphasized the desirability of particular comparisons between experimental information around the $^{132}$Sn and look for the corresponding nuclear properties in the $^{208}$Pb region. 

In semi-magic nuclei, the seniority quantum number distinguishes between the different nuclear states with identical valence nucleons occupying one orbital or a simple configuration according to the nucleon number  \cite{Racah}. This property provides us with an appropriate framework to establish a similarity between the $^{132}$Sn and $^{208}$Pb regions in terms of the orbital in each model-space using the shell-model wave-function. Seniority $\nu$ is the number of particles not in pairs coupled to angular momentum $J$ = 0. 
 In the seniority scheme, the properties of a nucleus can be described from the neighboring isotopes or isotones with  the addition or removal of valence nucleons \cite{Shalit,Talmi1,Talmi2,Shlomo}. 
Studies of even-even semi-magic nuclei within the generalized seniority scheme \cite{scholten,sandul,karta} have shown that the seniority-zero state can define the ground state. And the seniority $\nu > 0 $  states by pairs coupled to two unpaired nucleons represent low-lying excited states.

The $N$ = 82 region has long been the subject of shell-model studies due to the remarkable regularities exhibited in the low-lying states of these nuclei and also the strong doubly magic character of $^{132}$Sn. This region gives an opportunity over a large number of nuclei  ($A$ = 133 to 152) to study the effect of adding protons to a doubly magic core. Extensive research of this type was conducted over the past few years with the emergence of the shell when particles and  (or ) holes were added to the $^{132}$Sn \cite{Andreo1, Wildenthal1}. In this context, experimental data of energy levels and electromagnetic properties for light even-even $N$ = 82 isotones are rich, and have encouraged many shell-model calculations \cite{sarkar,Wildenthal2,Berant, Brown,Astier,Bianco}. Shell-model calculations using effective interactions derived from realistic nucleon-nucleon ($NN$) potentials have been performed in the $N$ = 82 region with some excellent results \cite{Andreo2, Covello1, holt, Suhonen, Covello2}. The evolution of the low-energy properties of the $N$ = 82 isotones as a function of $A$ has been well reproduced by the theory in some of these studies \cite{Covello1,holt,Suhonen}.  

The lead isotopes and the $N$=126 nuclei have long been the subject of both experimental and theoretical studies. These stable or near stable nuclei have been extensively investigated. A relatively large amount of experimental data is available in this region. While the high-spin states in nuclei near $^{208}$Pb have been studied extensively, data on electromagnetic transition rates between the low-spin states are insufficient. The good doubly magic character of $^{208}$Pb suggests successful shell model results and has motivated many shell-model calculations. On the other hand,
the heavy mass $^{208}$Pb region with protons and neutrons filling different major shells offers favorable conditions for the formation of isomeric states due to the presence of high-$j$ orbitals. With the close resemblance between proton-neutron multiplets in the $^{132}$Sn and $^{208}$Pb regions, an analogue of $^{208}$Pb \cite{Coraggio1, Coraggio2,Coraggio3,Coraggio4,Cie}, has led to a new focus on low-spin states in the lead region. The energy levels of $^{210}$Bi, $^{212}$Bi, $^{212}$At \cite{Coraggio1,Coraggio2,Coraggio3} were performed by employing an effective interaction derived from Bonn-A $NN$ interaction. Using the highlights of previous works by different groups, the structural information and the electromagnetic properties can be predicted with the shell-model calculation for the Sn and Pb region.

There are experimental indications for the similar properties of nuclei belonging to $N$ = 82 and $N$ = 126 regions, but no systematic theoretical results are available. Motivated with the recent experimental data for spectroscopic properties, we have performed a systematic shell model study of $N=82$ ($52 \leq Z \leq 60$) and $N=126$ ($84 \leq Z \leq 92$) isotones. The aim of the present paper is to investigate the similarity between $N=82$ and $N=126$ isotones. The present study will add more information to the earlier works. 

 These two isotonic chains are very rich in terms of nuclear isomers \cite{astier1,Jain,Stuchbery,Walker}. Thus we have also reported shell model results of nuclear isomers in these two regions. For the $N=82$ region we have discussed $6^+$ and $17/2^+$ isomers, while for 
 $N=126$ region the $8^+$, $11^-$, $21/2^-$ and $29/2^+$ isomers. The role of $g_{7/2}^2$ and $(g_{7/2}d_{5/2})^n$ configurations in the $N=82$ region, while  $h_{9/2}^n$, $(h_{9/2}f_{7/2})^n$ and $(h_{9/2}i_{13/2})^n$ configurations in $N=126$ are very crucial to decide nuclear isomers in these two regions.

This paper is organized as follows. In the Sec. \ref{formalism}, we present details about theoretical formalism. 
Comprehensive discussions are reported in Sec. \ref{results}. In this section, we have discussed spectroscopy of $N=82$ and $N=126$ isotones in terms of seniority. Finally, a summary and conclusions are drawn in Sec. \ref{summary}.

\section{Formalism: model space and interactions}
\label{formalism}

The shell-model calculations for the N = 82 isotones have been performed in the 50-82 valence shell composed of the orbits $1g_{7/2}$, $2d_{5/2}$, $2d_{3/2}$, $3s_{1/2}$, and  $1h_{11/2}$ using the interaction SN100PN taken from Brown et al.\cite{SN100PN1,Brown}. This interaction has four parts, corresponding to the neutron-neutron, neutron-proton, proton-proton component of the nuclear forces and Coulomb repulsion between the protons. The residual two-body interaction was obtained from the CD-Bonn G-matrix within the so-called $\hat{Q}$ -box folded diagram theory and the further nn part was multiplied by a factor of 0.90 to improve the agreement with the experiment for $^{130}$Sn. The single-particle energies for the neutrons are -10.610, -10.289, -8.717, -8.694, and -8.816 MeV for the $1g_{7/2}$, $2d_{5/2}$, $2d_{3/2}$, $3s_{1/2}$, and $1h_{11/2}$ orbitals, respectively, and those for the protons are 0.807, 1.562, 3.316, 3.224, and 3.605 MeV. We have performed calculations with the SN100PN interaction in some previous studies in this region \cite{2019, 2015, 2013}.

The shell-model calculations for the N = 126 isotones have been performed in the valence shell 82-126 for proton and 126-184 for neutron using the interaction KHPE \cite{Warburton1}. The model space here consists of $1h_{9/2}, 2f_{7/2}, 2f_{5/2}, 3p_{3/2}, 3p_{1/2}, 1i_{13/2}$ proton orbitals and  $1i_{11/2}, 2g_{9/2}, 2g_{7/2}, 3d_{5/2}, 3d_{3/2}, 4s_{1/2}, 1j_{15/2}$ neutron orbitals. The effective realistic residual interaction of Kuo and Herling \cite{Kuo1, Kuo2} was derived from a free nucleon-nucleon potential of Hamada and Johnston \cite{Hamada} with renormalization due to the finite extension of model space by the reaction matrix techniques developed by Kuo and Brown \cite{Kuo3}. In the present work, we have performed shell model calculations with KHPE interaction without any truncation. Recently, we have performed an extensive calculation for Rn isotopes with the KHPE interaction \cite{Rn}. In the present work, we have used the shell model code KShell \cite{kshell} for the diagonalization of matrices.

\section{Results and discussion} 
\label{results}

\subsection{$N$ = 82 even isotones}

\begin{figure*}[h]
 \includegraphics[width=\textwidth,height=\textheight,keepaspectratio]{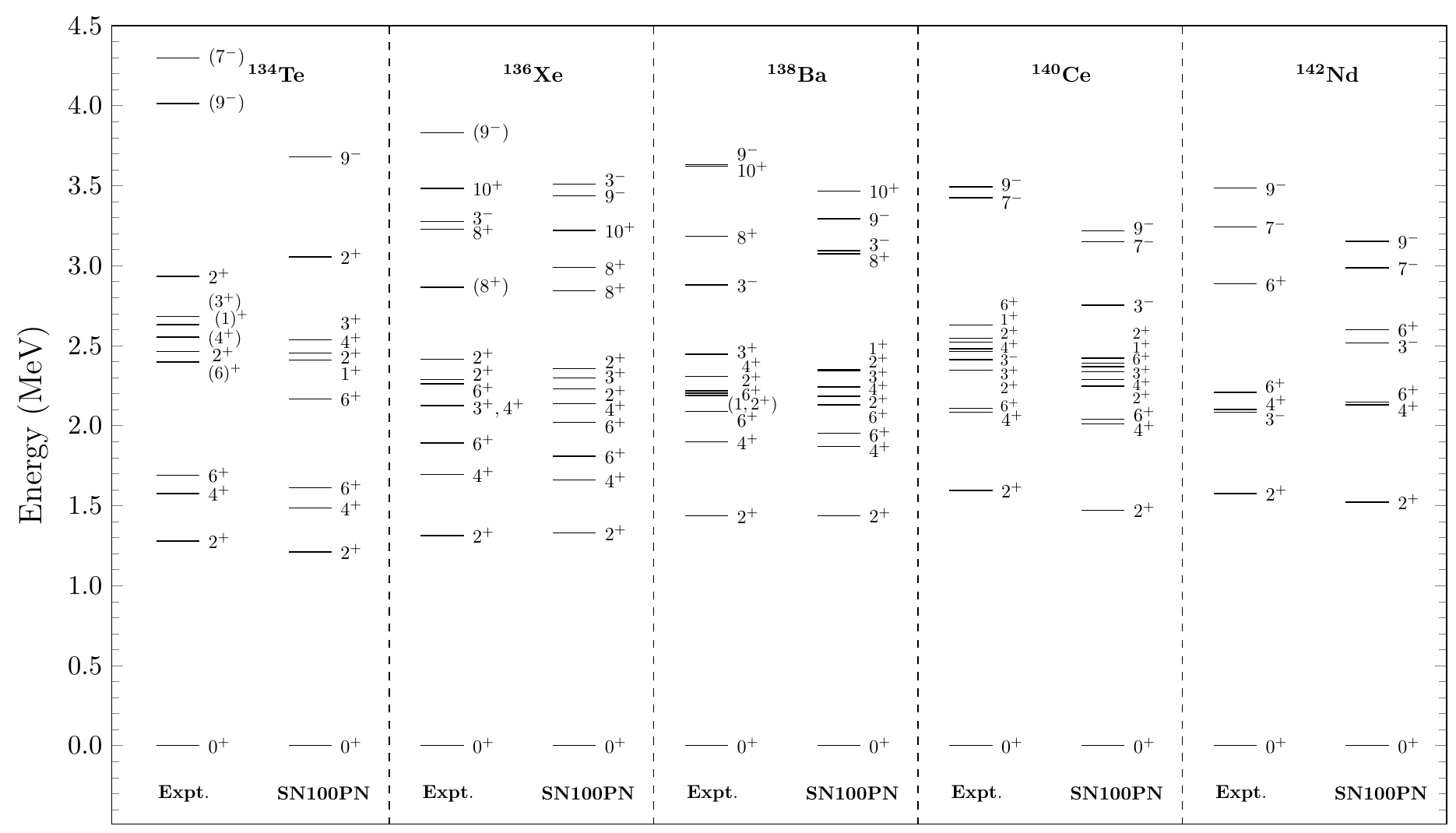}
  \caption{ Comparison of calculated and experimental \cite{nndc} excitation energy spectra for the $N$ = 82 even isotones using SN100PN interaction.}
  \label{even82}
\end{figure*}

\begin{figure*}[h]
 \includegraphics[width=\textwidth,height=\textheight,keepaspectratio]{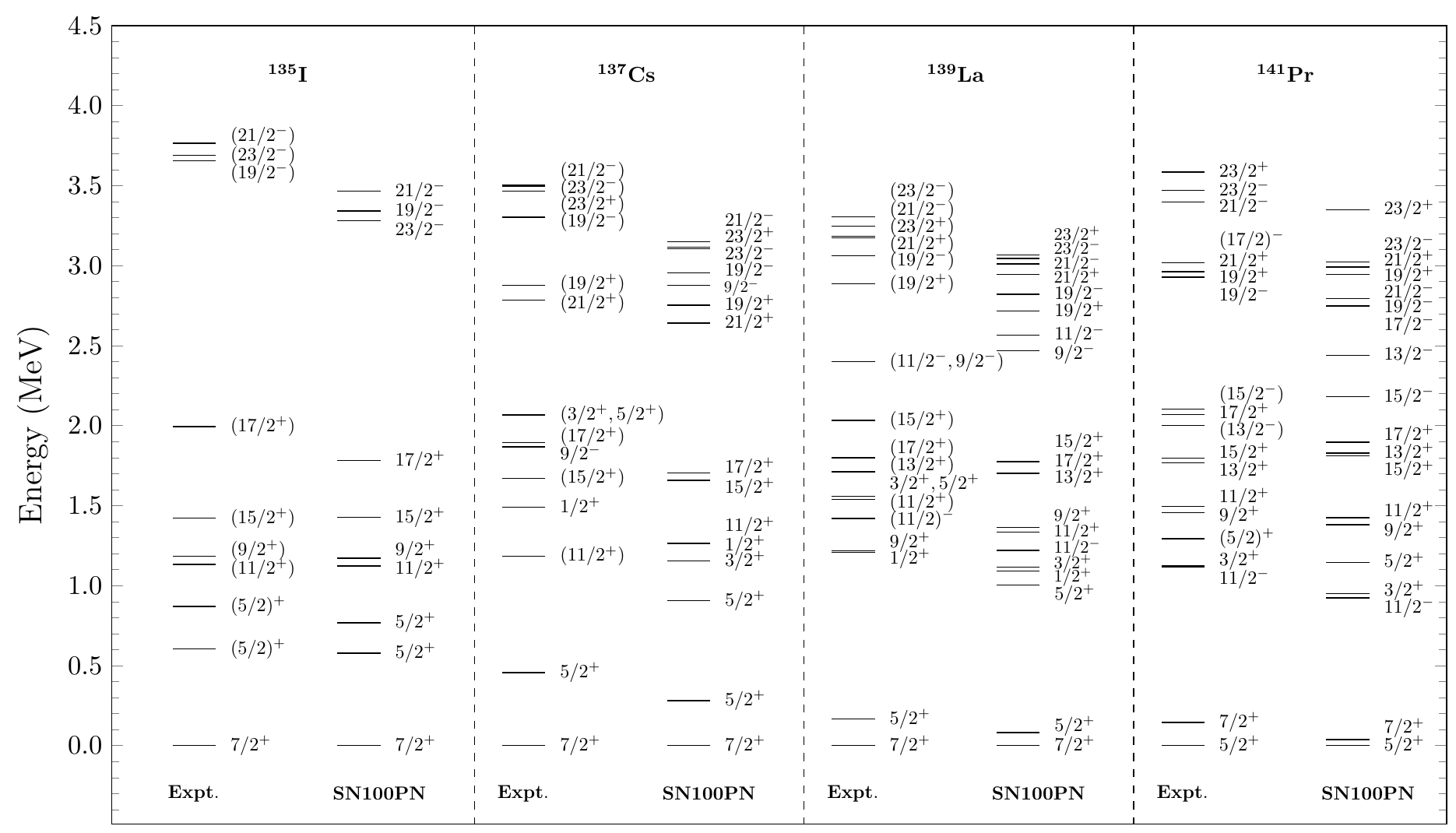}
\caption{\label{odd82} Comparison of calculated and experimental \cite{nndc} excitation energy spectra for the $N$ = 82 odd isotones using SN100PN interaction.}
\end{figure*}

In Figures \ref{even82}, \ref{odd82} we present the calculated energy levels of $N$ = 82 isotones for even and odd nuclei, respectively, obtained in the $50-82$ model space in comparison with the experimental data. In the $N$ = 82 isotones, the states correspond to proton excitations, for which the configurations are built within the proton orbitals ($1g_{7/2}$, $2d_{5/2}$, $2d_{3/2}$, $3s_{1/2}$, $1h_{11/2}$). The study of wave functions suggests that for all the low-lying positive-parity states, the valence protons occupy the two orbitals, $\pi g_{7/2}$ and $\pi d_{5/2}$. The main configurations of the low-lying positive-parity states correspond to $(\pi g_{7/2})^n$, $\pi (g_{7/2})^{n-1} d_{5/2}$, and $\pi (g_{7/2})^{n-2} d_{5/2}^2$ or configuration mixing of these components, where $n$ is the number of valence protons. We can simply represent these configurations as $\pi (g_{7/2} d_{5/2})^n$. The main structure of negative-parity states in the $N$ = 82 isotones above $^{132}$Sn is constructed by removing one proton from the $1g_{7/2}2d_{5/2}$ model space to the $\pi h_{11/2}$ orbital. Then the lowest negative-parity state gives rise to $(\pi g_{7/2} d_{5/2})^{n-1} h_{11/2}$ configuration. Our calculation shows that different configurations identify these low-lying states with varied natures. In our calculation, we take the low-lying yrast states, i.e., the assembling of the lowest state for each spin. One can see that the shell-model calculation very well reproduces the experimental energy levels. 

The $^{134}$Te contains two valence protons outside the $^{132}$Sn core. The present calculations reproduce the characteristic of spherical even-even nuclei that there is a large gap of about 1.2 MeV between ground state $0^+$ and the first excited state $2^+$ as shown in Fig \ref{even82}. The wave functions of the calculated states are pure, indicating the dominance of single-particle excitations in this nucleus. The lowest four states of $^{134}$Te arise from the $\pi g_{7/2}^2$ configuration. The ground state of $^{134}$Te may be described by two protons residing in the $\pi g_{7/2}$ orbital as one pair coupled to angular momentum $J$ = 0, i.e., seniority $\nu$ = 0. For the yrast chain of $\pi g_{7/2}^2$ configuration with seniority $\nu$ = 2, the $4^+_1$ and $6^+_1$ levels are also reproduced well. The other positive-parity states with energy between 2-3 MeV from proton excitation within the major shell are reproduced well. These states have the main configuration of $\pi g_{7/2}d_{5/2}$, with seniority $\nu$=2.
 The $2^+_3$ level at 3.055 MeV, which is close to the experimental $2^+$ level at 2.934 MeV, has a configuration of  $\pi d_{5/2}^2$ (71\%). 
The negative-parity states are observed above 3.5 MeV. These states belong to the multiplets of $\pi g_{7/2}h_{11/2}$, with seniority $\nu$=2. The levels of $9^-$, $7^-$ from the $\pi g_{7/2}h_{11/2}$ multiplet have good accordance with experimental data in both energy and order. 

The low-lying states in $^{136}$Xe can be well understood in terms of seniority, with a $N$ = 82 closed neutron shell and four protons beyond the $Z$ = 50 closed shell. The $\pi g_{7/2}^4$ configuration with seniorities $\nu$ = 0 and 2 gives rise to a state with $J^\pi$ = $0^+$ and a multiplet with $J^\pi$ = $2^+_1$, $4^+_1$, $6^+_1$, $4^+_2$, $2^+_3$, $8^+_1$  in order of increasing energy. These states are well reproduced, with over 50-90\% probability of this configuration in the wave functions of the respective calculated states. For the $0^+$ ground state of $^{136}$Xe this contribution amounts to 53\%. The wave functions of the states are rather pure with small admixtures, indicating the dominance of single-particle excitations.  Different multiplets arising from three protons in the $\pi g_{7/2}$ orbital and a single proton in the $\pi d_{5/2}$ orbital  ($\nu$ = 2 states) and two protons in each orbital ($\nu$ = 0 states) are also possible. The $\pi g_{7/2}^3 d_{5/2}$ $\nu$ = 2 configuration produces a $J^\pi$ = $2^+_2$, $3^+_1$, $6^+_2$, $8^+_2$, $10^+_1$ multiplet, where the even-spin states are reproduced well. The $6^+_2$ state at 2.262 MeV decays only to the $6^+_1$ state and not to the $4^+_1$ state, is observed in \cite{nndc}. 
The calculation shows that the states $9^-_1$, $3^-_1$ are the member of the $\pi g_{7/2}^3 h_{11/2}$ $\nu$ = 2 multiplet.

For $^{138}$Ba, the multiplets of $\pi g_{7/2}^4 d_{5/2}^2$ are $J^\pi$ = $0^+_1$, $2^+_1$, $2^+_2$, $4^+_1$, $6^+_2$. However, the contribution of $\pi g_{7/2}^6$  is also not small in the $0^+_1$ , $2^+_1$, $4^+_1$ and $6^+_2$ states,  with 20-30\% probability of this configuration. For the states $J^\pi$ = $6^+_1$, $4^+_2$, $3^+_1$, $2^+_3$, $1^+_1$,  $8^+_1$, $10^+_1$ the leading configuration is $\pi g_{7/2}^5 d_{5/2}$, contribution amounts to 40-60\%. The state $3^-_1$ is mainly due to the configuration $\pi g_{7/2}^4 d_{5/2}h_{11/2}$. The state $9^-_1$ is mainly due to one-proton excitation from the $\pi g_{7/2}$ orbital to $\pi h_{11/2}$, with the predominant configuration of $\pi g_{7/2}^5 h_{11/2}$. One can see the $0^+_1$ state has seniority $\nu$ = 0, and the even parity states as well as the $3^-_1$, $9^-_1$ states have seniority $\nu$ = 2. There are a few pair states which have two or three pairs, such as the $8^+_1$, $10^+_1$ states with seniority $\nu$=4. This indicates that the generalized seniority is a good quantum number here. 

$^{140}$Ce can be described similarly to $^{138}$Ba with two more protons in the valence space. The $\pi g_{7/2}^6 d_{5/2}^2$ $\nu$=0,2 configuration gives rise to  $J^\pi$ = $0^+_1$, $2^+_1$, $4^+_1$, $2^+_2$, $4^+_2$, $6^+_2$ multiplets. 
The $\pi g_{7/2}^5 d_{5/2}^3$ $\nu$=2 configuration produces $J^\pi$ = $6^+_1$, $5^+_1$, $3^+_1$, $1^+_1$, $2^+_3$, multiplet in order of increasing energy. However, these states from the two configuration chain are arising from significant configuration mixing with $\pi g_{7/2}^4 d_{5/2}^4$ and $\pi g_{7/2}^7 d_{5/2}$, respectively for the two chains, with 12-20\% probabilities of these configurations mixing with over 35-45\% of the leading configuration. Similar to $^{138}$Ba, the states $9^-_1$ and $7^-_1$ are mainly due to one-proton excitation from the $\pi g_{7/2}$ orbital to $\pi h_{11/2}$, with the predominant configuration of $\pi g_{7/2}^5 d_{5/2}^2 h_{11/2}$. The calculation shows that the state $3^-_1$ is a member of the $\pi g_{7/2}^6 d_{5/2} h_{11/2}$ multiplet.

For $^{142}$Nd, we compare selected calculated levels to their experimental counterparts. The $\pi g_{7/2}^6 d_{5/2}^4$ $\nu$=0,2 configuration gives rise to a $J^\pi$ = $0^+_1$, $2^+_1$, $4^+_1$, $6^+_2$ multiplet. However, these states are arising from significant configuration mixing with $\pi g_{7/2}^6 d_{5/2}^2 h_{11/2}^2$ with 10-15\% probability of this configuration mixing with over 20-35\% of the leading configuration. Contrast to $^{138}$Ba and $^{140}$Ce, the states $3^-_1$ and $7^-_1$ are mainly due to one-proton excitation from the $\pi d_{5/2}$ orbital to $\pi h_{11/2}$ and the $9^-_1$ state is due to one-proton excitation from the $\pi g_{7/2}$ orbital to $\pi h_{11/2}$ with the predominant configuration of $\pi g_{7/2}^6 d_{5/2}^3 h_{11/2}$ and $\pi g_{7/2}^5 d_{5/2}^4 h_{11/2}$, respectively. Thus these negative parity states are seniority $\nu$ = 2 states.

Interestingly, it can be seen that the $9^-_1$ isomeric state of $^{134}$Te and $^{136}$Xe is better represented by coupling the octupole phonon to the $6^+_2$ non-isomeric state rather than to the $6^+_1$ isomeric state, while the $9^-_1$ isomeric state of $^{138}$Ba, $^{140}$Ce and $^{142}$Nd, is better represented by coupling the octupole phonon to the $6^+_1$ isomeric state. Particularly, it was suggested for $^{136}$Xe and $^{138}$Ba in Ref \cite{Hwang}. From this calculation,  a broad understanding into the structure of the low-lying levels can be obtained.

\subsection{$N$ = 82 odd isotones}

$^{135}$I is a single-closed nucleus with three valence protons. The present results well explain those measured experimental energies of the low-lying positive-parity states. Most of the experimental states in $^{135}$I are tentative, and our calculation confirms all the proposed states with spin-parity in correct order. In $^{135}$I, the low-lying energies are obtained mainly in terms of the three configurations, $\pi g_{7/2}^3$, $\pi g_{7/2}^2 d_{5/2}$, and $\pi g_{7/2}^2 h_{11/2}$, which are mostly pure states with dominant configuration probability over 75-100$\%$. The lowest-energy configuration $\pi g_{7/2}^3$ can create  $J^\pi$ = $5/2^+$, $7/2^+$, $9/2^+$, $11/2^+$ and $15/2^+$. Similar to $J^\pi$ = $0^+$ with seniority $\nu$ = 0 in $^{134}$Te, the ground state $J^\pi$ = $7/2^+$ with seniority single-proton, $\nu$ = 1 in $^{135}$I. Our calculated wave functions indicate that the $5/2^+_2$, $7/2^+_1$, $9/2^+_1$, $11/2^+_1$ and $15/2^+_1$ states have the predominant configuration $\pi g_{7/2}^3$. 
The $\nu$ = 3 multiplet contains the $5/2^+_2$ , $9/2^+_1$, $11/2^+_1$ and $15/2^+_1$ states where the corresponding compositions are over 90$\%$.
These states, therefore correspond to the single-particle nature. The dominant configuration of the lower $5/2^+_1$ state is $\pi g_{7/2}^2 d_{5/2}$ ($\nu$ = 1). The $17/2^+_1$ state is found to be the member of the same multiplet as $5/2^+_1$ state, but $\nu$ = 3.  The negative-parity states $19/2^-_1$, $23/2^-_1$, and $21/2^-_1$ between 3 and 4 MeV are due to one proton excitation from the orbital $\pi g_{7/2}$ into the orbital $\pi h_{11/2}$. The predominant configuration is $\pi g_{7/2}^2 h_{11/2}$, with seniority $\nu$ = 3.

For $^{137}$Cs, the calculated low-lying yrast states are due to configuration mixing. The $J^\pi$ = $7/2^+_1$, $5/2^+_2$, $3/2^+_1$, $11/2^+_1$, and $15/2^+_1$ states have a leading configuration of $\pi g_{7/2}^5$, but with some differences. The $7/2^+_1$ and $11/2^+_1$ states are mixed with a considerable amount of $\pi g_{7/2}^3 d_{5/2}^2$ component, while this component is small in $15/2^+_1$. The $7/2^+_1$ ground state, in which two $0^+$ proton pairs are coupled to the odd $g_{7/2}$ proton is seniority $\nu$ = 1 state. As in $^{135}$I, the $\nu$ = 3 states are $J^\pi$ = $5/2^+_2$, $3/2^+_1$, $11/2^+_1$, and $15/2^+_1$, excitations with a $\pi g_{7/2}$ broken pair coupled to a proton in the $\pi g_{7/2}$. In $^{137}$Cs an extra pair of valence protons is available to generate  high-spin yrast states. The $5/2^+_1$, $1/2^+_1$, $17/2^+_1$, and $19/2^+_1$ states belong to the $\pi g_{7/2}^4 d_{5/2}$ multiplet with seniority $\nu$ = 3, with a $\pi g_{7/2}$ broken pair coupled to a proton in the $\pi d_{5/2}$, except for $21/2^+_1$ which is seniority $\nu$ = 5 state. The $23/2^+_1$ is a member of the $\pi g_{7/2}^3 d_{5/2}^2$, $\nu$ = 5. The negative parity excitations involve one proton from  $\pi g_{7/2}$ to $\pi h_{11/2}$ orbital, with $\nu$ = 3. The negative parity states $J^\pi$ = $9/2^-_1$, $19/2^-_1$, $21/2^-_1$ and $23/2^-_1$ are multiplets of $\pi g_{7/2}^4 h_{11/2}$ configuration.

For $^{139}$La, the calculated low-lying yrast states for $J$= $1/2$ to $23/2$ are due to the mixing of several configurations, except for $19/2^+_1$, $21/2^+_1$ and $23/2^+_1$, where leading configuration dominates. 
The $J^\pi$ = $7/2^+_1$, $5/2^+_2$, $11/2^+_1$, $15/2^+_1$, $19/2^+_1$, and $23/2^+_1$ states have a leading configuration of $\pi g_{7/2}^5 d_{5/2}^2$, configuration probability ranges from 39-82$\%$. The $1/2^+_1$ state is arising from $\pi g_{7/2}^4 d_{5/2}^2 s_{1/2}$ configuration with 24\% probability. The $7/2^+_1$ ground state and $1/2^+_1$ are seniority $\nu$ = 1 state, in which $0^+$ proton pairs are coupled to the odd $g_{7/2}$ and $s_{1/2}$ proton, respectively. The $\nu$ = 3 states with $J^\pi$ = $5/2^+_2$, $11/2^+_1$, $15/2^+_1$ and $\nu$ = 5 states $19/2^+_1$, $23/2^+_1$, are arising due to coupling of  $\pi g_{7/2}$ or $\pi d_{5/2}$ broken pair (or pairs) with  $\pi g_{7/2}$. The $\pi g_{7/2}^4 d_{5/2}^3$ seniority $\nu$ = 3 and $\nu$ = 5, gives rise to the multiplet $3/2^+_1$, and $21/2^+_1$, respectively. The $J^\pi$ = $5/2^+_1$, $9/2^+_1$, $13/2^+_1$ and $17/2^+_1$ states have a leading configuration of $\pi g_{7/2}^6 d_{5/2}$, excitations with a $\pi g_{7/2}$ broken pair coupled to a proton in the $\pi d_{5/2}$. The negative-parity states are due to one proton excitation from the orbital $\pi g_{7/2}$ or $\pi d_{5/2}$ into the orbital $\pi h_{11/2}$. The dominant configuration $\pi g_{7/2}^4 d_{5/2}^2 h_{11/2}$, with seniority $\nu$ = 3 gives rise to the multiplet $9/2^-_1$, $11/2^-_1$ and $23/2^-_1$. The $11/2^-_2$ and $21/2^-_1$ are $\nu$ = 3 states, with dominant configuration $\pi g_{7/2}^5 d_{5/2} h_{11/2}$. The $19/2^-_1$ state is a $\nu$ = 3 state with dominant configuration $\pi g_{7/2}^6 h_{11/2}$ (41$\%$), and has a significant contribution from $\pi g_{7/2}^4 d_{5/2}^2 h_{11/2}$ (34$\%$) configuration.

For $^{141}$Pr, the excitation energies and ordering of the positive-parity yrast states are well reproduced in the present calculations. The negative parity states are with some discrepancies, as they are compressed in energy with respect to the experimental data. The $^{141}$Pr has similar multiplet structure as $^{139}$La, as well the calculated wave functions have a significant admixture of configurations involving excitations of the $\pi g_{7/2}$ or $\pi d_{5/2}$ orbital. Compared to low-lying states, relatively less configuration mixing is observed at high-lying states. Whereas, in $^{141}$Pr, the number of multiplets and maximum spin achievable is large with an extra pair of valence protons in $\pi g_{7/2}$ or $\pi d_{5/2}$ orbital. In $^{141}$Pr, $5/2^+$ state becomes the ground state as the number of protons increases, in which $0^+$ proton pairs are coupled to the odd $d_{5/2}$ is a seniority $\nu$ = 1 state, with configuration $\pi g_{7/2}^6 d_{5/2}^3$. Similar to $^{139}$La, the $\pi g_{7/2}^6 d_{5/2}^3$ multiplets are seniority $\nu$ = 3 or 5, with a $\pi g_{7/2}$ broken pair(s) coupled to a proton in the $\pi d_{5/2}$, with $J^\pi$ = $3/2^+_1$, $9/2^+_1$, $17/2^+_1$, $19/2^+_1$ and $21/2^+_1$, where $19/2^+_1$ and $21/2^+_1$ are $\nu$ = 5 states. The $\pi g_{7/2}^5 d_{5/2}^4$ multiplets are $J^\pi$ = $5/2^+_2$, $7/2^+_1$, $11/2^+_1$, $15/2^+_1$ with $\nu$ = 3 and $23/2^+_1$ is $\nu$ = 5 state. The $1/2^+_1$ is a $\nu$ =1 state of $\pi g_{7/2}^6 d_{5/2}^2 s_{1/2}$ configuration, and $13/2^+_1$ is a $\nu$ =3 state of $\pi g_{7/2}^7 d_{5/2}^2$ configuration. As in $^{139}$La with two added protons, the negative-parity states are mainly due to $\pi g_{7/2}^5 d_{5/2}^3 h_{11/2}$ and $\pi g_{7/2}^6 d_{5/2}^2 h_{11/2}$ configurations  with seniority $\nu$ = 3. The first group has members $J^\pi$ = $17/2^-_1$, $19/2^-_1$, $21/2^-_1$ and $23/2^-_1$ states, the states with $J^\pi$ = $11/2^-_1$, $13/2^-_1$ and $15/2^-_1$ are member of second group.

\subsection{$N$ = 126 even isotones}
\begin{figure*}
 \includegraphics[width=\textwidth,height=\textheight,keepaspectratio]{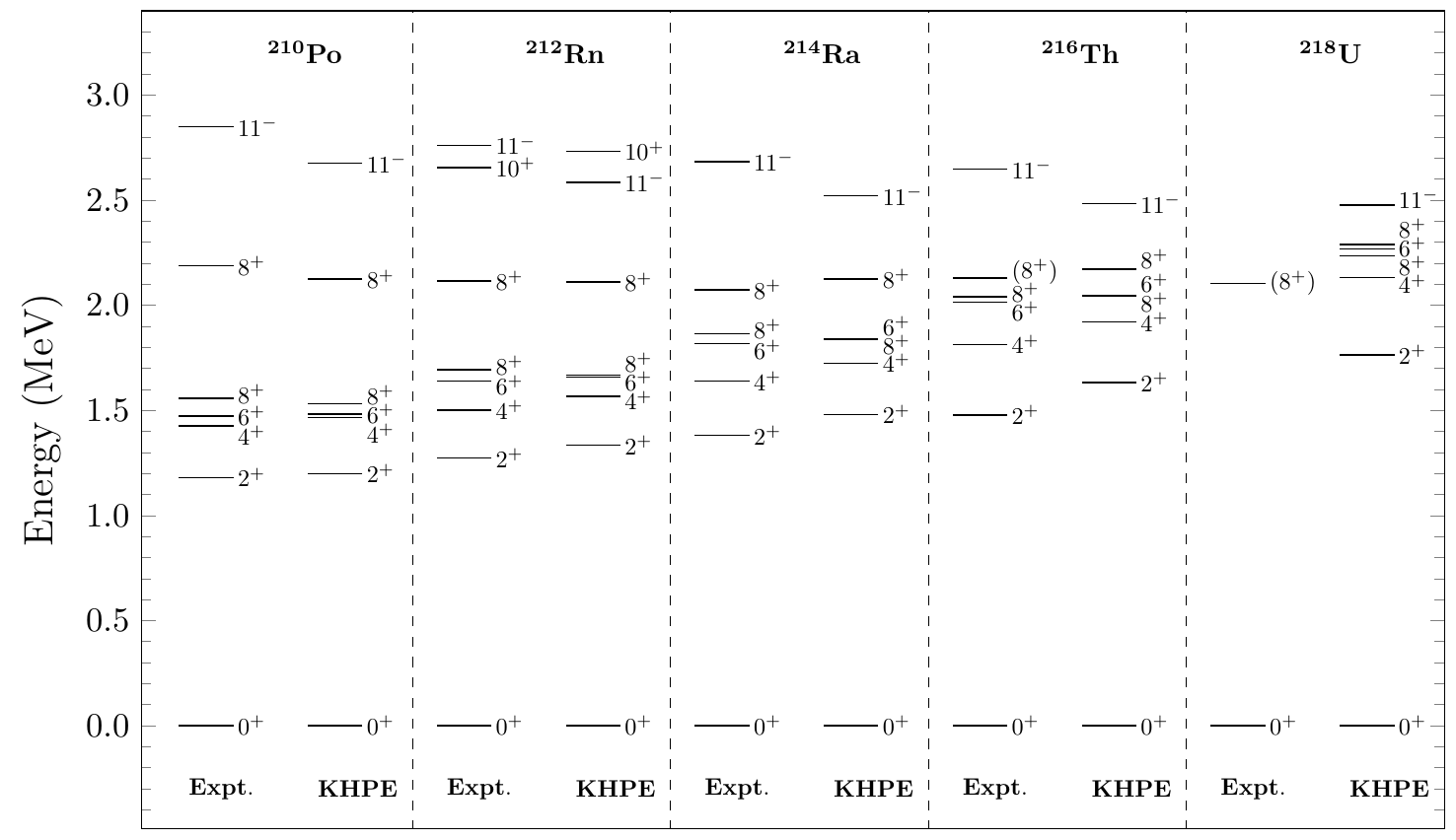}
  \caption{\label{even126} Comparison of calculated and experimental \cite{nndc} excitation energy spectra for the $N$ = 126 even isotones using KHPE interaction.}
\end{figure*}

Isotones with $N$ = 126 represent an excellent example of the applicability of the seniority scheme with $\nu \ge 2$, as the low-lying spectrum follows the seniority scheme for the semi-magic nuclei above the $^{208}$Pb core. Calculations in the entire range of proton configurations using KHPE interaction showed a good agreement with the experimental data. This calculation confirmed the seniority conservation for the isotones up to $Z$ = 92. 

\begin{figure*}
 \includegraphics[width=\textwidth,height=\textheight,keepaspectratio]{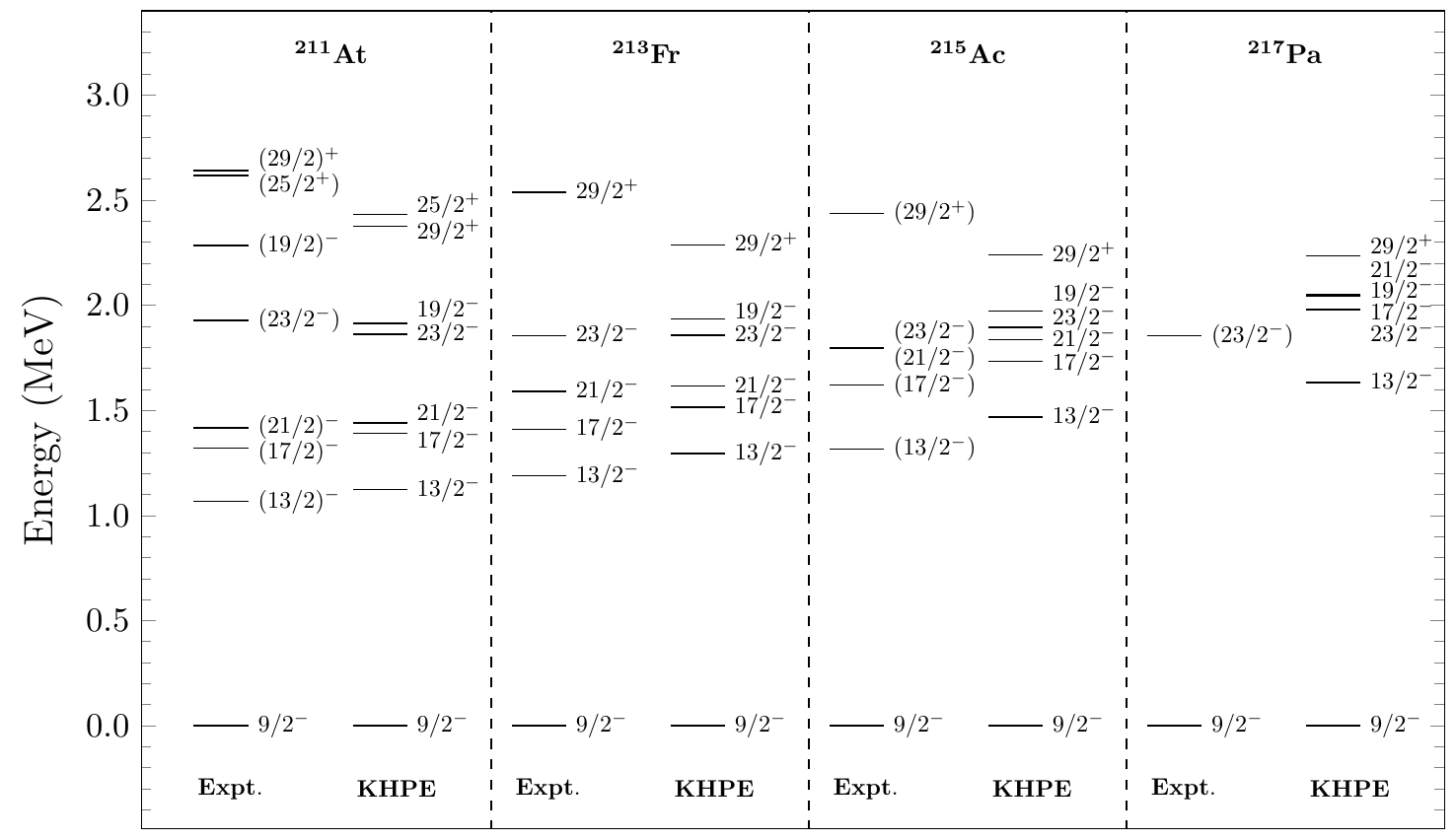}
 \caption{\label{odd126} Comparison of calculated and experimental \cite{nndc} excitation energy spectra for the $N$ = 126 odd isotones using KHPE interaction.}
\end{figure*}

In Figures \ref{even126}, \ref{odd126} we present the calculated energy levels of $N$ = 126 isotones for even and odd nuclei, respectively, obtained in the shell-model configuration space in comparison with the experimental data. In the $N$ = 126 isotones, the low-lying states correspond to proton excitations, for which the configurations are built within the proton orbitals ($1h_{9/2}, 2f_{7/2}, 2f_{5/2}, 3p_{3/2}, 3p_{1/2}, 1i_{13/2}$). The study of wave functions suggests that for all the low-lying positive-parity states, the valence protons occupy the three orbitals, $\pi h_{9/2}$, $\pi f_{7/2}$ and $\pi i_{13/2}$. The main configurations of the low-lying positive-parity states correspond to $(\pi h_{9/2})^n$, $\pi (h_{9/2})^{n-1}  f_{7/2}$, $\pi (h_{9/2})^{n-2} f_{7/2}^2$, and $\pi (h_{9/2})^{n-4} f_{7/2}^2 i_{13/2}^2$ or combinations of them, where $n$ is the number of valence protons. We can simply represent these configurations as $(\pi h_{9/2} f_{7/2} i_{13/2})^n$. The main structure of negative-parity states in the $N$ = 126 isotones above $^{208}$Pb is simply constructed by exciting one proton from the $\pi h_{9/2} f_{7/2}$ model space to the $\pi i_{13/2}$ orbital, and also from $\pi (h_{9/2} f_{7/2})^n$ for odd isotones. Our calculation shows that these low-lying states are identified by different configurations with a combination of mixed and pure states. In our calculation, we take the low-lying yrast states. One can see the experimental energy levels  are very well reproduced for the $N$ = 126 isotones by the shell-model calculation.

$^{210}$Po contains two valence protons outside the $^{208}$Pb core. The low-lying yrast states, $0^+_1$, $2^+_1$, $4^+_1$, $6^+_1$ and $8^+_1$ are constructed by proton-dominated two-particle excitations of the $\pi h_{9/2}^2$ configuration, in a similar fashion to the situation in $^{134}$Te. The wave functions of the states are pure, indicating the dominant single-particle excitations in $^{210}$Po. In contrast, two protons need to occupy the $\pi i_{13/2}$ orbital to make states above $10^+$. The $0^+_1$ ground state is a seniority $\nu$ = 0 state. The states $J^\pi$ = $2^+_1$, $4^+_1$, $6^+_1$ and $8^+_1$ are seniority $\nu$ = 2 states, with one broken proton pair. The $8^+_2$ state is a member of the $\pi h_{9/2} f_{7/2}$ configuration and seniority $\nu$ = 2. The $11^-_1$ state consists of the $\pi h_{9/2} i_{13/2}$ configuration. 

The low-lying states in $^{212}$Rn are made up of four valence protons beyond the $Z$ = 82 closed shell, with a $N$ = 126 closed neutron shell should be well understood in terms of seniority structure, same as $^{136}$Xe. In the first four excited states $J^\pi$ = $2^+_1$, $4^+_1$, $6^+_1$ and $8^+_1$ the percentage of the dominant $\pi h_{9/2}^4$ configurations, ranges from 60\% to 65\% while it becomes 42\% in the ground state. All the levels arising from the $\pi h_{9/2}^4$ configuration have a configuration mixing with the $\pi h_{9/2}^2 i_{13/2}^2$ and $\pi h_{9/2}^2 f_{7/2}^2$ configurations. 
The ground state of $^{212}$Rn may be described by four protons residing in the $\pi h_{9/2}$ orbital as two pairs coupled to angular momentum $J$ = 0, i.e., seniority $\nu$ = 0. The excited states are seniority $\nu$ = 2 states of one broken pair with the dominant $\pi h_{9/2}^4$ configuration. The $8^+_2$ excited state dominated by the $\pi h_{9/2}^3 f_{7/2}$ configuration is a seniority $\nu$ = 2 state with one odd $\pi h_{9/2}$ proton coupled to $\pi f_{7/2}$, with  71\% contribution. The $8^+_2$ state has a significant mixing with $\pi h_{9/2} f_{7/2} i_{13/2}^2$ and $\pi h_{9/2} f_{7/2}^3$ configurations. The $10^+_1$ state is dominated by $\pi h_{9/2}^4$ configuration, with over 95\% contribution. The $10^+_1$ is seniority $\nu$ =4 state with the breaking of two proton pairs. The low-lying $11^-$ dominated by $\pi h_{9/2}^3 i_{13/2}$ configuration is a seniority $\nu$ = 2 state.

For $^{214}$Ra, $^{216}$Th and $^{218}$U, the low-lying states are due to significant configuration mixing. In $^{214}$Ra, the components of $\pi h_{9/2}^6$ are $J^\pi$ = $2^+_1$, $4^+_1$, $6^+_1$, $8^+_1$. These states are arising from significant configuration mixing with $\pi h_{9/2}^4 f_{7/2}^2$ and $\pi h_{9/2}^4 i_{13/2}^2$, with 24\% probabilities of these configurations mixing with over 28-32\% of the leading configuration. The ground state of $^{214}$Ra can be described by $\pi h_{9/2}^4 i_{13/2}^2$ with 24\% probability as two pairs coupled to angular momentum $J$ = 0, i.e., seniority $\nu$ = 0. However, the $\pi h_{9/2}^6$ (21\%) and $\pi h_{9/2}^4 f_{7/2}^2$ (21\%) components are also not small for the ground state. The $8^+_2$ excited state dominated by the $\pi h_{9/2}^5 f_{7/2}$ configuration is a seniority $\nu$ = 2 state with two proton pairs and one odd $\pi h_{9/2}$ proton coupled to $\pi f_{7/2}$, with over 43\% contribution. The state $11^-_1$ is mainly due to one-proton excitation from the $\pi h_{9/2}$ orbital to $\pi i_{13/2}$, with the predominant configuration of $\pi h_{9/2}^5 i_{13/2}$. One can see the $0^+_1$ state is seniority $\nu$ = 0 state, and the even parity states as well as the $11^-_1$  are seniority $\nu$ = 2 states.

$^{216}$Th can be described similarly to $^{214}$Ra with two more protons in the valence space. The $\pi h_{9/2}^6 f_{7/2}^2$, $\nu$=2 configuration gives rise to a $J^\pi$ = $2^+_1$, $4^+_1$, $6^+_1$, $8^+_1$ multiplets. However, these states are arising from significant configuration mixing with $\pi h_{9/2}^6 i_{13/2}^2$ and $\pi h_{9/2}^4 f_{7/2}^2 i_{13/2}^2$, with over 16-20\% probabilities of these configurations mixing with over 20-22\% of the leading configuration. Similar to $^{214}$Ra with two extra proton in $\pi f_{7/2}$ and $\pi i_{13/2}$ orbital, the ground state can be described by seniority $\nu$ = 0 with $\pi h_{9/2}^4 f_{7/2}^2 i_{13/2}^2$ configuration and the $8^+_2$ is a seniority $\nu$ = 2 state dominated by the $\pi h_{9/2}^5 f_{7/2} i_{13/2}^2$ configuration. The $0^+_1$ is due to admixture of $\pi h_{9/2}^6 f_{7/2}^2$ and $\pi h_{9/2}^6 i_{13/2}^2$ configurations with almost the same contribution 16-17\% as the dominant one (18\%). The $11^-_1$ is a seniority $\nu$ = 2 state of $\pi h_{9/2}^5 i_{13/2}^3$ configuration with 23\%, including the admixture of $\pi h_{9/2}^5 f_{7/2}^2 i_{13/2}$ and $\pi h_{9/2}^7 i_{13/2}$ configurations with over 20-22\% contribution.

Showing similar structure as $^{216}$Th the multiplets from the configuration $\pi h_{9/2}^6 f_{7/2}^2$  including the ground state now arises from $\pi h_{9/2}^6 f_{7/2}^2 i_{13/2}^2$ in $^{218}$U with two added valence protons. The multiplets $J^\pi$ = $0^+_1$, $2^+_1$, $4^+_1$, $6^+_1$, $8^+_2$ are due to significant mixing with several configurations, with 8-9\% probabilities of these configurations mixing with 19-22\% of the leading configuration. In $^{218}$U the contribution of configurations apart from the dominant one has decreased compared to $^{214}$Ra and $^{216}$Th. The seniority scheme here is the same as the $^{214}$Ra and $^{216}$Th. In contrast to other even $N$ = 126 isotones, the $8^+_1$ state is a multiplet of $\pi h_{9/2}^5 f_{7/2}^3 i_{13/2}^2$, highly mixed with $\pi h_{9/2}^7 f_{7/2}^3$ and $\pi h_{9/2}^7 f_{7/2} i_{13/2}^2$ configurations with almost  the  same contribution 13-15\% as the dominant one (over 15\%). The $11^-_1$ is a multiplet of $\pi h_{9/2}^5 f_{7/2}^2 i_{13/2}^3$ with 19\%, including the admixture of $\pi h_{9/2}^7 f_{7/2}^2 i_{13/2}$ and $\pi h_{9/2}^7 i_{13/2}^3$ configurations with over 16\% contributions.

\subsection{$N$ = 126 odd isotones}

The spectra of the odd-even isotones calculated assuming the pure $h_{9/2}^n$ configuration with seniority $\nu$ = 1 and 3 are compared with the experimental data. The experimental spectra themselves no longer contain all the levels predicted by the seniority scheme. These isotones belong to the region of neutron-rich short-lived nuclei and, the precise determination of their spectra of excited states is a complex experimental task.

The low-lying levels of $^{211}$At can be described by the coupling of three valence protons residing outside the $Z$ = 82, $N$ = 126 closed shell. The level scheme of $^{211}$At, $^{213}$Fr and $^{215}$Ac are very similar.
In $^{211}$At, all the spins considered are pure states with 96-100\% contribution and for ground state over 67\% contribution. The $^{210}$Po and $^{211}$At both are confined to a single orbital $\pi h_{9/2}$ for the low-lying states. The results are explained relating in a $\pi h_{9/2}^2$ configuration ($^{210}$Po) to those in a $\pi h_{9/2}^3$ configuration ($^{211}$At). The $J^\pi$ = $9/2^-_1$, $13/2^-_1$, $17/2^-_1$ and $21/2^-_1$ states are identically defined by three protons in $\pi h_{9/2}$ with seniority $\nu$ = 3, except for $9/2^-_1$. For $9/2^-_1$, there are two independent states with seniority $\nu$ = 1 and 3, respectively. In $^{211}$At, the $9/2^-_1$ is the ground state and we can assume it has seniority $\nu$ = 1. The $19/2^-_1$ and $23/2^-_1$ states are formed by $\pi h_{9/2}^2 f_{7/2}$ configuration with seniority $\nu$ = 3. The $29/2^+_1$ and $25/2^+_1$ are seniority $\nu$ = 3 states with one proton excitation from $\pi h_{9/2}$ to $\pi i_{13/2}$ orbital, which is $\pi h_{9/2}^2 i_{13/2}$ configuration.  The energy of the $19/2^-_1$ state is higher compared to other first negative parity states. Because the single-particle energy of the $\pi f_{7/2}$ orbital is higher than that of the $\pi h_{9/2}$ orbital, the energy of the $19/2^-_1$ state is higher compared to the $17/2^-_1$ and $21/2^-_1$ states. The $17/2^-_1$ and $21/2^-_1$ states consist of the $\pi h_{9/2}^3$ configuration. In contrast the $19/2^-_1$ state cannot be constructed by the $\pi h_{9/2}^3$ configuration.

 In $^{213}$Fr, the calculated low-lying yrast states are due to configuration mixing. The $J^\pi$ = $9/2^-_1$, $13/2^-_1$, $17/2^-_1$ and $21/2^-_1$ states have a leading configuration of $\pi h_{9/2}^5$, with 38-60\% probability which is the  lowest for the ground state. The $9/2^-_1$ ground state, in which two $0^+$ proton pairs are coupled to the odd $h_{9/2}$ proton, is seniority $\nu$ = 1 state. As in $^{211}$At, the seniority $\nu$ = 3 states are $J^\pi$ = $13/2^-_1$, $17/2^-_1$ and $21/2^-_1$, excitations with a broken pair coupled to a proton in the $\pi h_{9/2}$ orbital. The $19/2^-_1$ and $23/2^-_1$ states belong to the $\pi h_{9/2}^4 f_{7/2}$ multiplet is seniority $\nu$ = 3 state, with a broken pair coupled to a proton in the $\pi f_{7/2}$ orbital. 
The $29/2^+_1$ is a seniority $\nu$ = 3 state having a leading configuration of $\pi h_{9/2}^4 i_{13/2}$, where two protons in the $\pi h_{9/2}$ orbital with spin eight coupled to the last proton in the $\pi i_{13/2}$ orbital. The maximum spin of this configuration is 29/2. 

For $^{215}$Ac, the calculated yrast states are due to large configuration mixing. The ground state $9/2^-_1$ has the leading configuration $\pi h_{9/2}^5 i_{13/2}^2$, although another configuration $\pi h_{9/2}^5 f_{7/2}^2$ has a significant probability almost the same as the leading term of 22\%. This shows the importance of $\pi i_{13/2}$ orbital for the formation of low-lying states around $Z$ = 90 and above it. The $J^\pi$ = $13/2^-_1$, $17/2^-_1$ and $21/2^-_1$ states have a leading configuration of $\pi h_{9/2}^5 f_{7/2}^2$, with probability over 27$\%$. However, these states are coming from the mixing of two more configurations, $\pi h_{9/2}^5 i_{13/2}^2$ and $\pi h_{9/2}^7$, with over 19-24\% contribution. The $23/2^-_1$ and $19/2^-_1$ states belong to the $\pi h_{9/2}^6 f_{7/2}$ multiplet, and the $29/2^+_1$ state  belongs to the $\pi h_{9/2}^6 i_{13/2}$ multiplet, with over 38\% contribution. The seniority scheme for the states followed here is the same as those in $^{211}$At and $^{213}$Fr.

For $^{217}$Pa, the calculated yrast states are also due to large configuration mixing. The $J^\pi$ = $9/2^-_1$, $13/2^-_1$ and $17/2^-_1$ states have a leading configuration of $\pi h_{9/2}^5 f_{7/2}^2 i_{13/2}^2$, with probability from 20-23$\%$,  which is the lowest for the ground state. Similar to the $^{215}$Ac, these states are coming from the mixing of two more configurations $\pi h_{9/2}^7 i_{13/2}^2$ and $\pi h_{9/2}^7 f_{7/2}^2$, with over 12-18\% contribution with the leading term. The $23/2^-_1$, $19/2^-_1$ and $21/2^-_1$ states belong to the $\pi h_{9/2}^6 f_{7/2} i_{13/2}^2$ multiplet with seniority $\nu$ = 3, with a $\pi h_{9/2}$ broken pair coupled to a proton in the $\pi f_{7/2}$. The $29/2^+_1$ state belongs to the $\pi h_{9/2}^6 f_{7/2}^2 i_{13/2}$ multiplet with seniority $\nu$ = 3, where a broken pair in the $\pi h_{9/2}$ orbital coupled to the last proton in the $\pi i_{13/2}$ orbital. 
In contrast to the other odd $N$ = 126 isotones, the excitation energy of the isomeric state $23/2^-_1$ is lower than the  $21/2^-_1$ ($\pi h_{9/2}^3$) at $Z$ = 91. Previously, this was also expected in the experimental work \cite{217pa}.

\subsection{Similarity between Sn and Pb regions}

The low-lying states are expected to rise only from the three high-$j$ orbitals $\pi g_{7/2}$, $\pi d_{5/2}$ and $\pi h_{11/2}$ for the $N$ = 82 isotones with $Z \ge$ 50. 
For the $N$ = 82 isotones ($52 \leq Z \leq 60$), the breaking of a first proton pair generates different multiplets occupying the $\pi g_{7/2}$ and $\pi d_{5/2}$ orbitals depending on the number of valence protons. 
In doubly even nuclei, the low-lying excited states $J^\pi \leq 6^+$ are generated from breaking of one proton pair with seniority $\nu \leq 2$ from the $\pi( g_{7/2}d_{5/2})^n$ configuration. The positive parity states with $J^\pi \ge 6^+$ are possible in the same configuration space coexisting with core-couple states. Hence, for the states, $J^\pi \ge 6^+$ seniority must be $\nu \ge 4$ with breaking a second or more proton pair (up to $J \leq 12$). Although the admixture of configuration from the $\pi d_{3/2}$ and $\pi s_{1/2}$  orbitals are small but very crucial to generate the correct level scheme \cite{140Ce, 141Pr}. In doubly even nuclei, odd parity states $J^\pi \leq 9^-$ are generated by the ($\pi h_{11/2}, \pi d_{5/2}$) and ($\pi h_{11/2}, \pi g_{7/2}$) multiplets.

\begin{figure*}
\begin{center}
  \includegraphics[width=70mm,height=65mm]{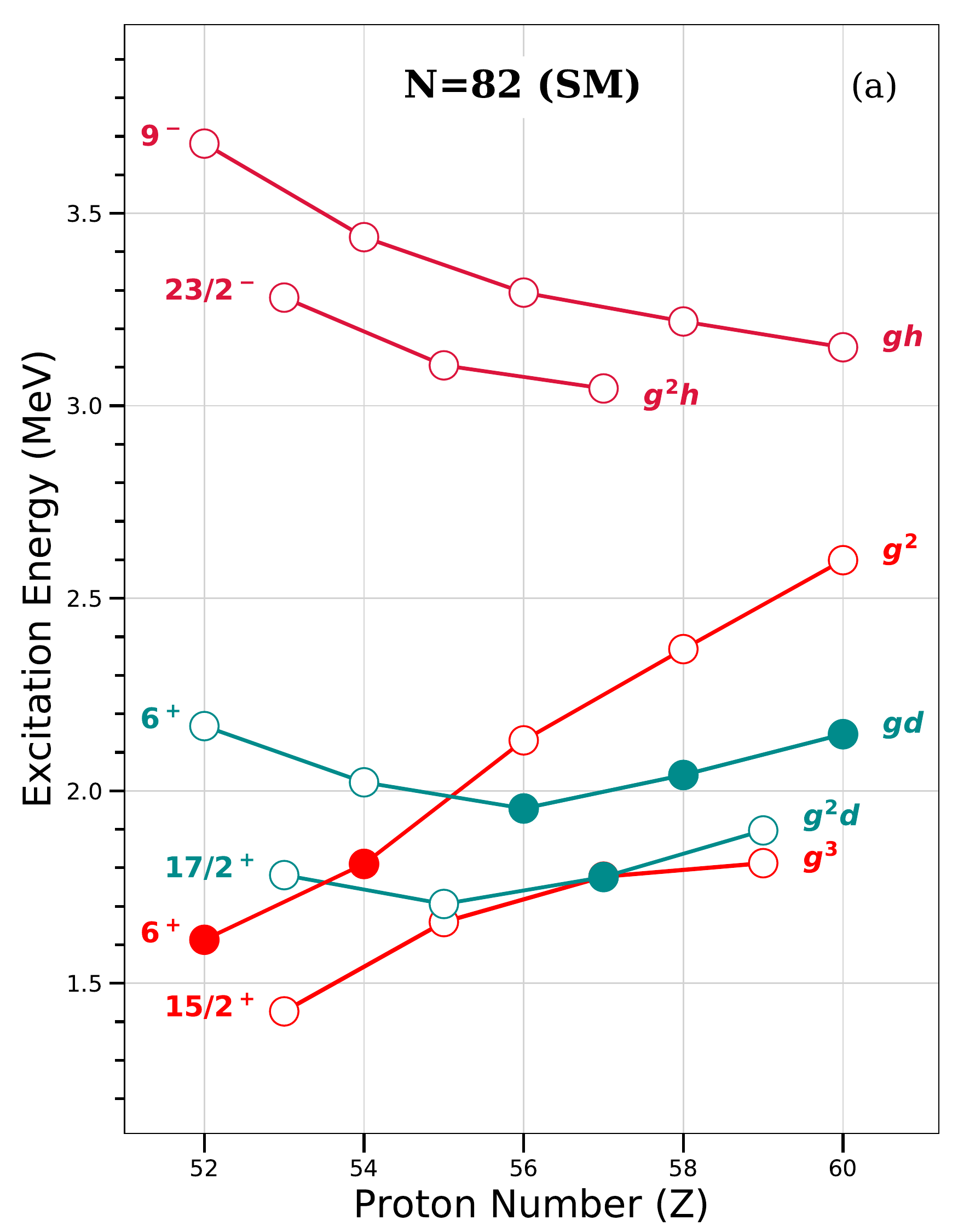}
  \includegraphics[width=70mm,height=65mm]{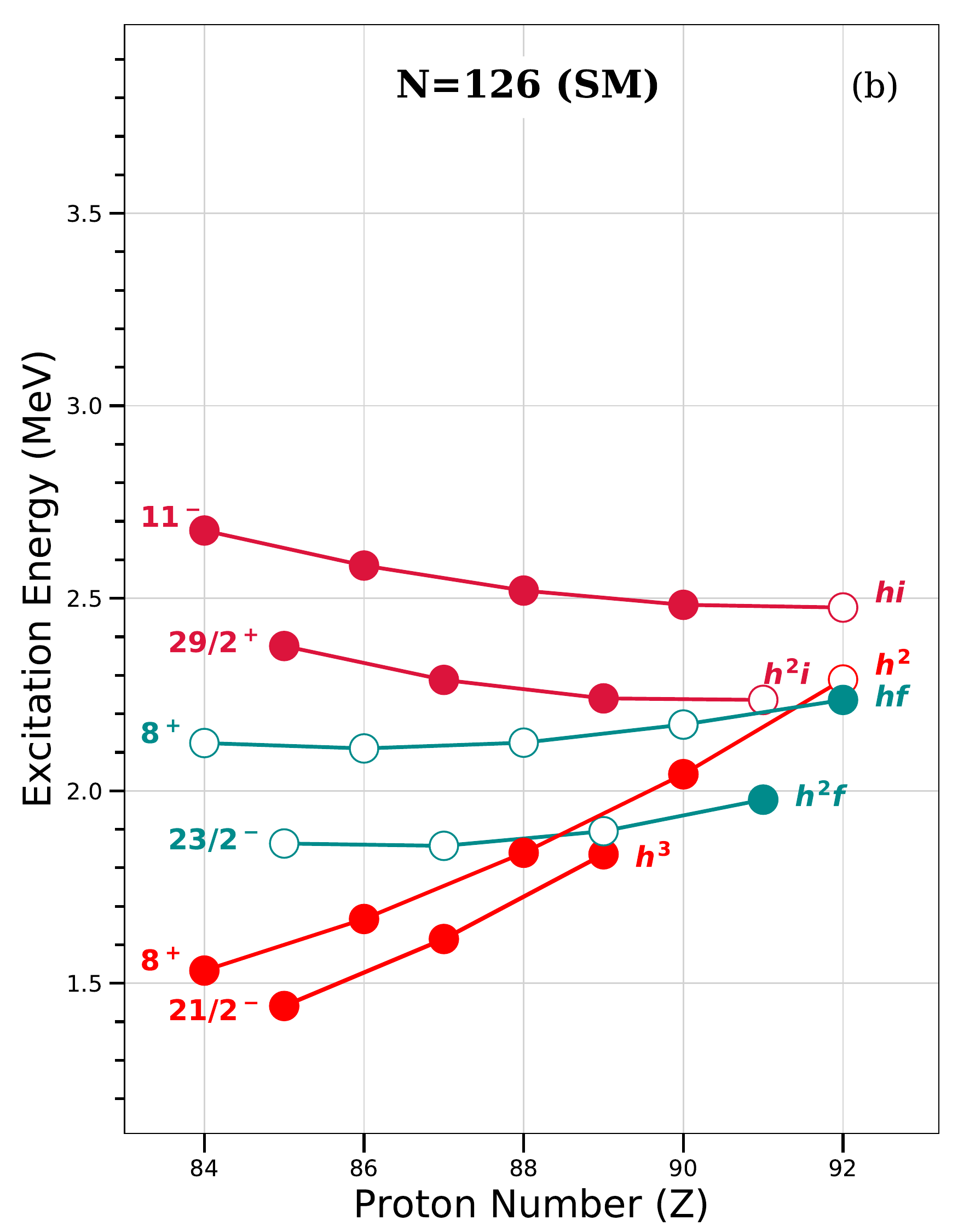}
\includegraphics[width=70mm,height=65mm]{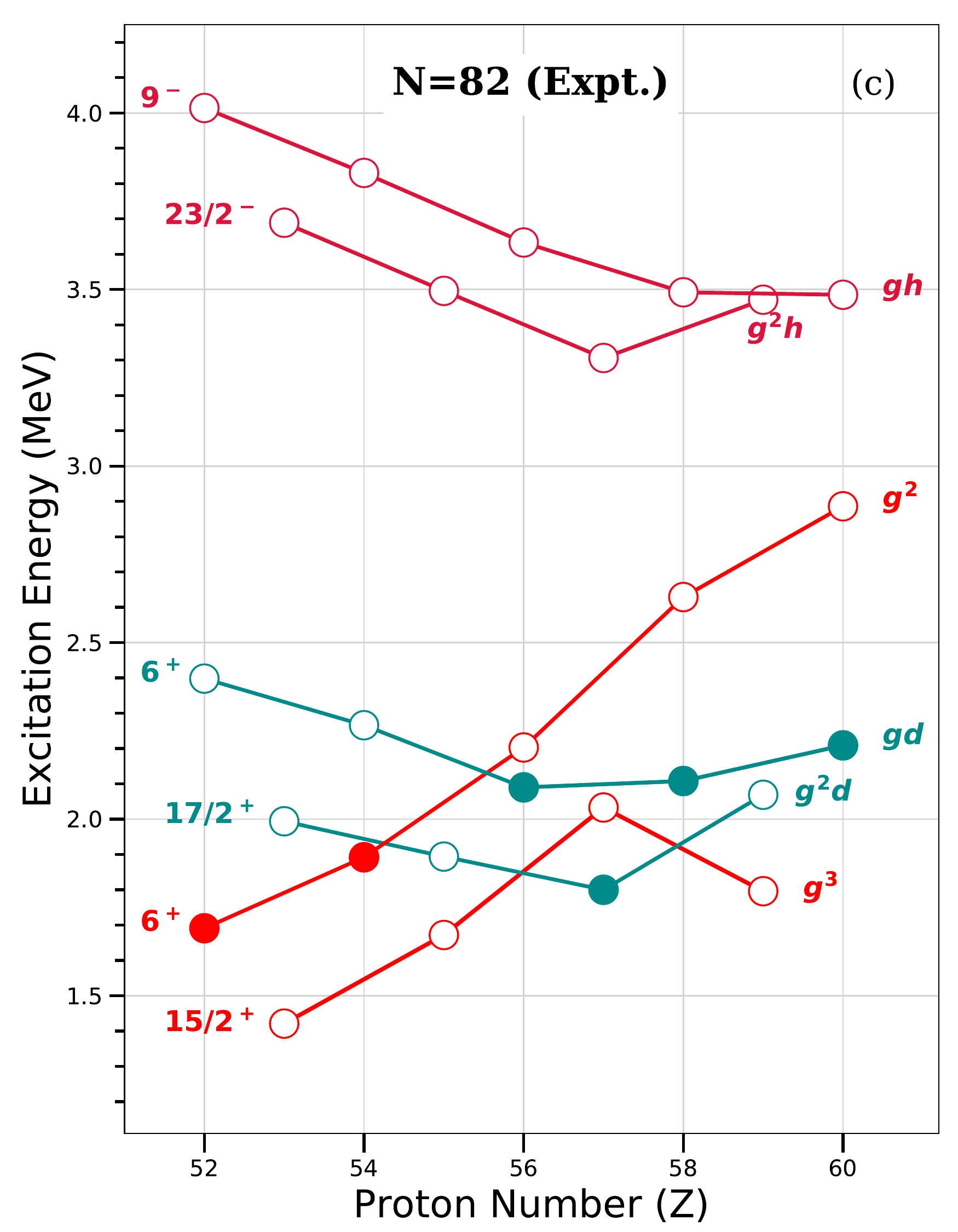}
  \includegraphics[width=70mm,height=65mm]{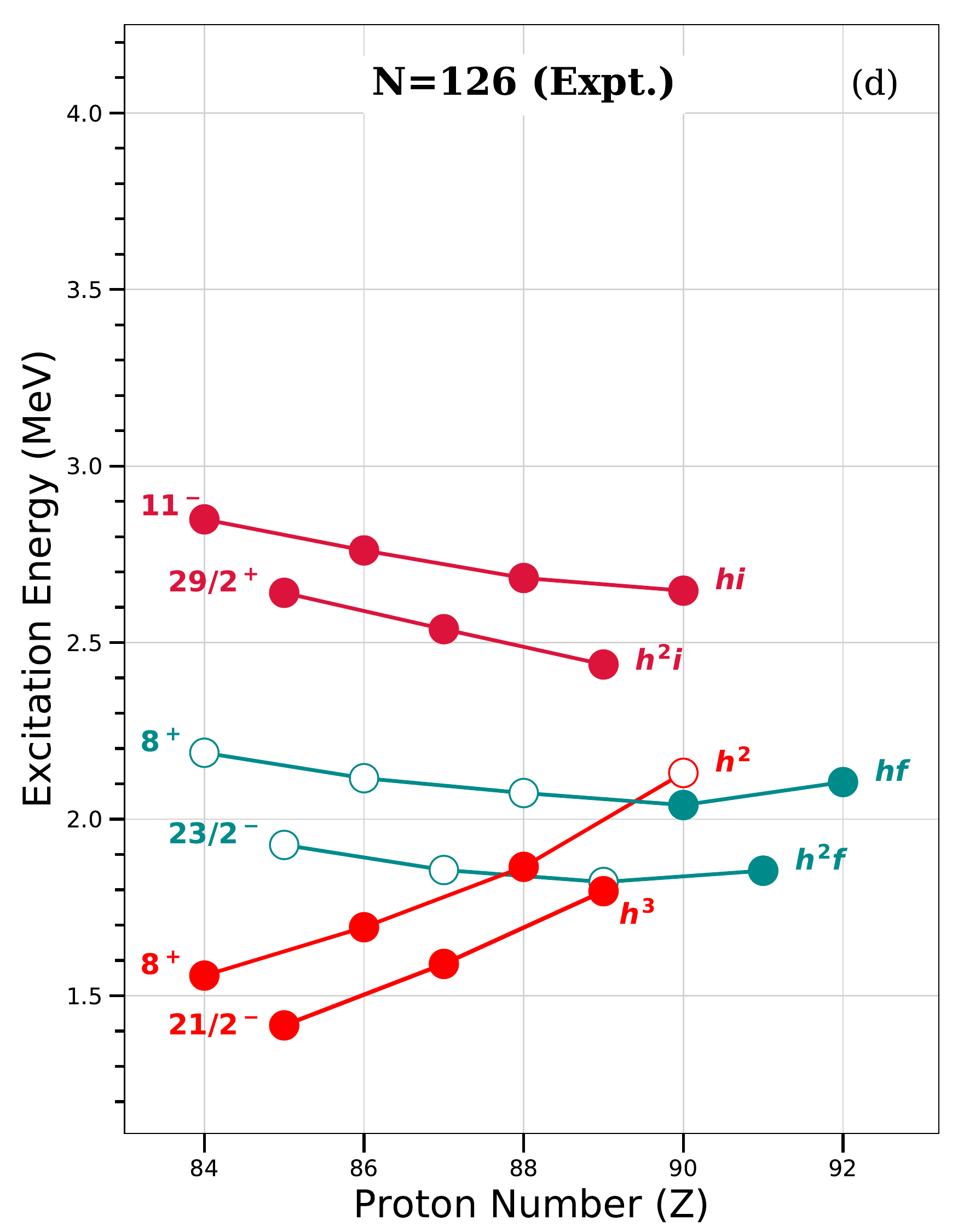}
  \label{fig:126exp}
\caption{\label{orbitals}Calculated excitation energies of the fully-aligned states with one broken proton pair in the $N$ = 82 isotones with $52 \leq Z \leq 60$  
   (a), and in the $N$ = 126 isotones with $84 \leq Z \leq 92$ (b) in comparison with experimental excitation energies for $N$ = 82 isotones (c), and $N$ = 126 isotones (d). The isomeric states are highlighted with filled symbols. }
\end{center}
\end{figure*}

\begin{figure*}
\begin{center}
     \includegraphics[width=73mm,height=70mm]{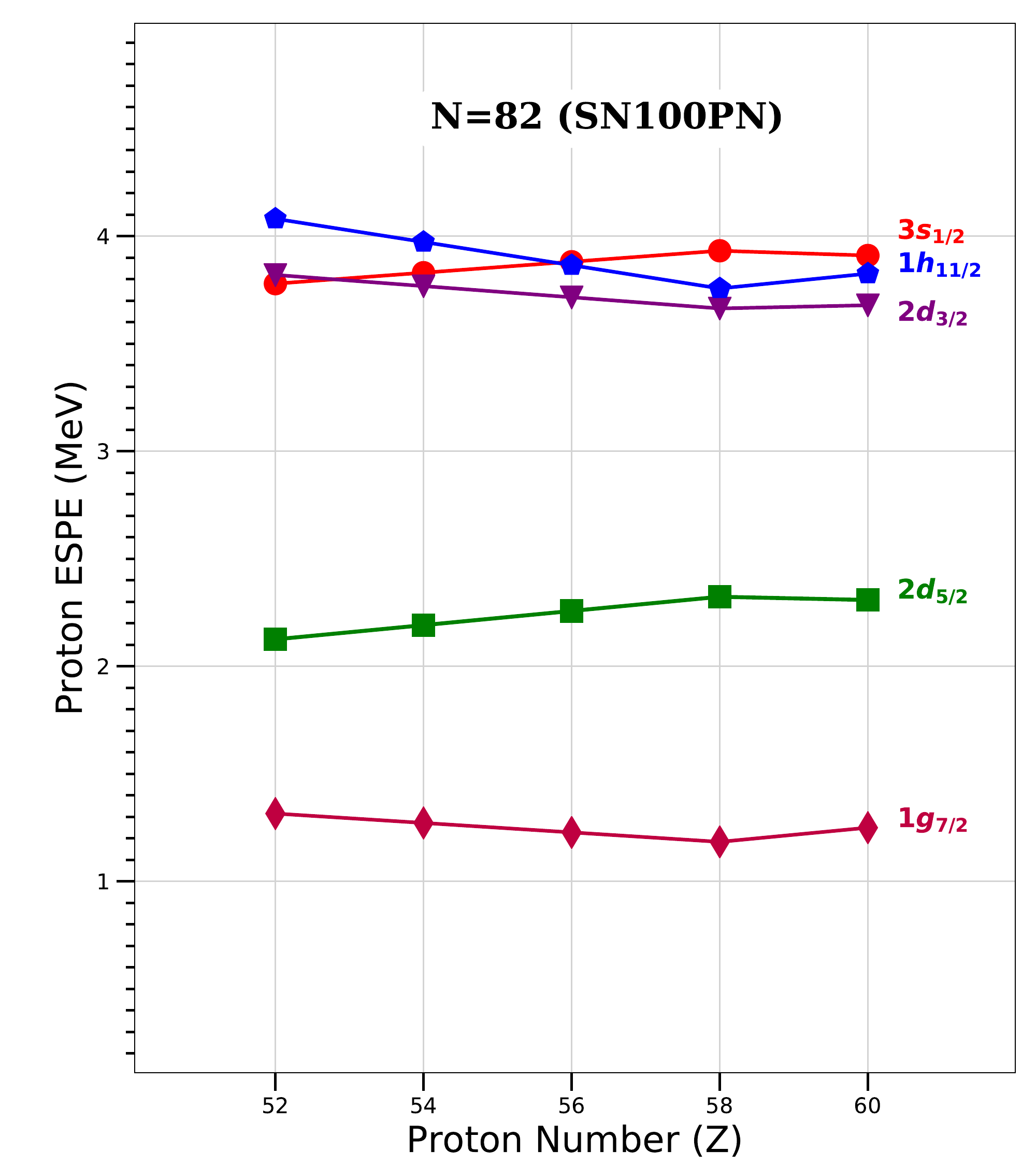}
     \includegraphics[width=73mm,height=70mm]{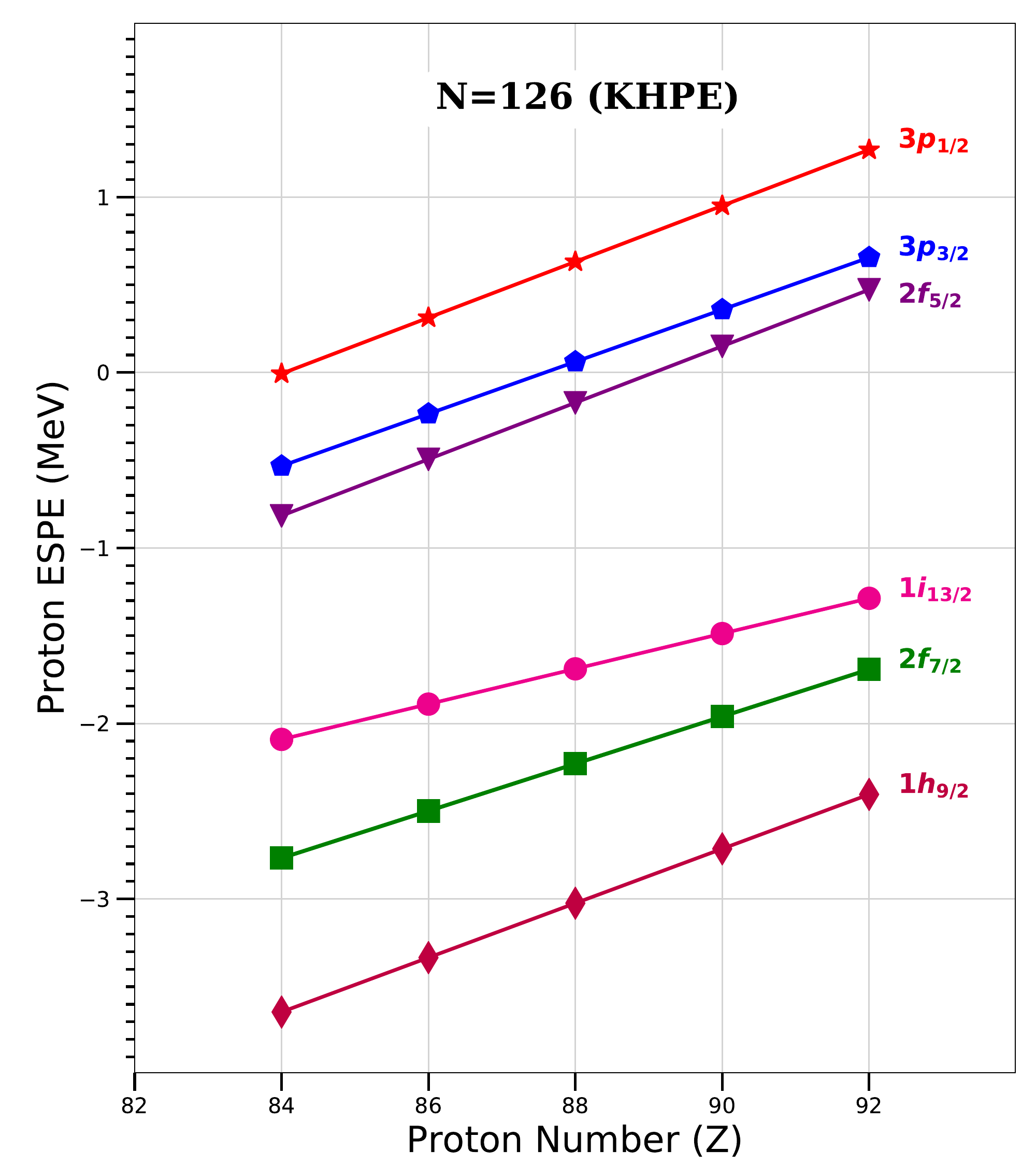}
  \label{fig:monopole}
\caption{\label{monopole} Variation of proton effective single-particle energies  for different orbitals of $N$ = 82 and $N$ = 126 isotones. }
\end{center}
\end{figure*}

Similarly, in the Pb-region, particular orbitals are consistently responsible for forming the low-lying states and isomers. The low-lying states are expected to rise only from the three high-$j$ orbitals $\pi h_{9/2}$, $\pi f_{7/2}$ and $\pi i_{13/2}$ for the $N$ = 126 isotones with $Z \ge$ = 82. For the $N$ = 126 isotones ($84 \leq Z \leq 92$), the breaking of a first proton pair generates different multiplets occupying the $\pi h_{9/2}$ and $\pi f_{7/2}$ orbitals depending on the number of valence protons.
In doubly even nuclei, the low-lying excited states $J^\pi \leq 8^+$ are generated from breaking of one proton pair with seniority $\nu \leq 2$ from the $\pi( h_{9/2}f_{7/2})^n$ configuration. The positive parity states with $J^\pi \ge 8^+$ are possible in the same configuration space coexisting with core-couple states. Hence, for the states $J^\pi \ge 8^+$ seniority must be $\nu \ge 4$ with breaking of a second or more proton pair (up to $J \leq 14$). In doubly even nuclei, odd parity states $J^\pi \leq 11^-$ are generated by the ($\pi i_{13/2}, \pi h_{9/2}$) and ($\pi i_{13/2}, \pi f_{7/2}$) multiplets, with seniority $1 \leq \nu \leq 4$  for states up to $J^\pi \leq 17^-$.

Fig. \ref{orbitals} shows the similarity between the shell structures in the $^{132}$Sn and $^{208}$Pb regions for the fully-aligned states with one broken proton pair in the $N$ = 82 isotones with $52 \leq Z \leq 60$, involving the three orbitals just above the $Z$ = 50 gap, $\pi g_{7/2}, \pi d_{5/2}$, and $\pi h_{11/2}$, and in the $N$ = 126 isotones with $84 \leq Z \leq 92$, involving the three orbitals just above the $Z$ = 82 gap, $\pi h_{9/2}, \pi f_{7/2}$, and $\pi i_{13/2}$. In these two regions the orbital angular momentum differ by one unit, with the same orientation of the intrinsic spin, $\pi g_{7/2} \rightarrow \pi h_{9/2}, \pi d_{5/2} \rightarrow \pi f_{7/2}$, and $\pi h_{11/2} \rightarrow \pi i_{13/2}$. Similar evolution can be seen for the same seniority $\nu$ = 2 states in the two $6^+$ states of the $N$ = 82 isotones and the two $8^+$ states of the $N$ = 126 isotones. Similarly for the odd isotones with seniority $\nu$ =3 the behavior between the $15/2^+$ and $17/2^+$ states of the $N$ = 82 isotones and the $21/2^-$ and $23/2^-$ states of the $N$ = 126 isotones.

 In Fig. \ref{monopole},  we have shown the variation of proton effective single-particle energies for $N$ = 82 and $N$ = 126 isotones corresponding to SN100PN and KHPE interactions, respectively. 
For $N=82$ isotones, as we increase proton number, the gap between proton effective single-particle energies for $\pi g_{7/2}$ and $\pi h_{11/2}$ orbital decreases. 
This effect can be seen as a decrease in energy (see Fig. \ref{orbitals} ) for $9^-$ and $23/2^-$ states having configurations composed of $\pi g_{7/2}$ and $\pi h_{11/2}$. 
For $N$ = 126 isotones with KHPE interaction, the gap between proton effective single-particle energies for $\pi h_{9/2}$ and $\pi i_{13/2}$ orbital decreases by 436 keV as we move from $Z=84$ to  $Z=92$, this is also reflected from the decrease in the energies of $11^-$ and $29/2^+$ isomers.

\begin{table*}[h]
		\caption{The calculated (SM) half-life with configuration for $6^+$ state in Sn-region and $8^+$ state Pb- region with respect to the experimental data (Expt.) \cite{nndc}.}
		\label{t_hl}
	\begin{ruledtabular}
		\begin{tabular}{c|ccccc|c|ccccc}
			&   &  &   & & &  &  & & &  &  \\
			$N$ = 82 & $\pi g_{7/2}^2$  & $\pi g_{7/2} d_{5/2}$  & Transition &  T$_{1/2}$  & T$_{1/2}$ & $N$ = 126 & $\pi h_{9/2}^2$ & $\pi h_{9/2} f_{7/2}$ & Transition & T$_{1/2}$ & T$_{1/2}$  \\
			 &   &    & & (SM)  & (Expt.) &  &  & & & (SM) & (Expt.)  \\
			&   &  &   & & &  &  &  & && \\
			\hline                                                                 

{$^{134}$Te} & $6^+_1$  &  -  & $E2$ & 126.90 ns  & 164.1(9) ns & {$^{210}$Po} & $8^+_1$ & - & $E2$ & 106.16 ns &  98.9(25) ns \\
& - & $6^+_2$ &  $M1+E2$ & 1.17 ps & $<$ 16 ps &  & - & $8^+_2$ & $M1+E2$ & 33.70 ps &  -   \\\hline

{$^{136}$Xe} & $6^+_1$  &  -  & $E2$ & 1.47 $\mu$s  & 2.95(17) $\mu$s & {$^{212}$Rn} & $8^+_1$ &  - & $E2$ & 1.83 $\mu$s  & 0.91(3) $\mu$s \\
& - & $6^+_2$ & $M1+E2$ & 10.72 ps & $\le$ 50 ps &  & - & $8^+_2$ & $M1$ & 37.56 ps &  -  \\\hline

{$^{138}$Ba} & -  &  $6^+_1$  & $E2$ & 3.60 $\mu$s  & 0.85(16) $\mu$s & {$^{214}$Ra} & $8^+_1$ & - & $E2$ & 130.88 $\mu$s &  67.8(15) $\mu$s \\
& $6^+_2$ & - & $M1+E2$ & 11.48 ps & 55(17) ps &  & - & $8^+_2$ & $M1$ & 58.23 ps &  -   \\\hline

{$^{140}$Ce} & -  &  $6^+_1$  & $E2$ & 13.95 $\mu$s  & 7.31(15) $\mu$s & {$^{216}$Th} & $8^+_1$ & - & $E2$ &  41.90 $\mu$s & 134 (4) $\mu$s   \\
& $6^+_2$ & - & $M1+E2$ & 6.26 ps &  & & - & $8^+_2$ & $M1$ & 25.16 ps &  -   \\\hline
			 
{$^{142}$Nd} & -  &  $6^+_1$  & $E2$ & 5279 $\mu$s  & 16.5 $\mu$s & {$^{218}$U} & - & $8^+_1$ & $E2$ & 26.39 $\mu$s & 0.56$^{+26}_{-14}$ ms  \\
 & $6^+_2$ & - & $M1+E2$ & 3.94 ps & &  & $8^+_2$ & - & $M1$ & 17.90 ps &   -  

		\end{tabular}
		\end{ruledtabular}
\end{table*}

It should be noted that the configuration $(\pi g_{7/2}d_{5/2})^n$ is more appropriate than taking the two orbitals individually as the two orbitals are close in energy. Similarly, for the configuration $(\pi h_{9/2} f_{7/2})^n$ in the Pb-region. Since we have two different orbitals in the configurations of the even isotones, two different combinations of orbitals can give the same angular momentum. Such as, $J^\pi = 6^+$ from  $\pi g_{7/2}^2$ and $\pi g_{7/2} d_{5/2}$ in Sn-region, and $J^\pi = 8^+$ from  $\pi h_{9/2}^2$ and $\pi h_{9/2} f_{7/2}$ in Pb-region. It is found that the low-lying $6^+_1$ is an isomeric state, regardless of the configuration. 

The different natures and the evolution of the two $6^+$ and $8^+$ states can be confirmed by the different values of the calculated half-lives in Table \ref{t_hl}. We can see the exchange of configuration in both the Sn-region (between two $6^+$) and Pb-region (between two $8^+$) at almost in the middle of the range of isotones $52 \leq Z \leq 60$ for $N$ = 82 and $84 \leq Z \leq 92$ for $N$ = 126. This is because of the small gap in the single-particle energy between the orbitals $\pi g_{9/2}$ and $\pi d_{5/2}$ in the Sn-region and $\pi h_{9/2}$ and $\pi f_{7/2}$ in the Pb-region. Although, from the shell-model calculation, the configuration exchange for $8^+$ is taking place at $Z$ = 92. The half-lives of $6^+_1$ and $8^+_1$ states are always larger than $6^+_2$ and $8^+_2$ states, confirming the isomeric nature for the first yarst states of these spins, despite the change of configuration. The $6^+_2$ state decays with  the $M1+E2$ transition, and the $M1$ transition shows a large impact by reducing the half-life from ns to ps. Similarly, the $8^+_2$ state is solely dominated by  the  $M1$ transition, thus giving a smaller half-life than the $8^+_1$ state. 
Our calculation supports the experimental counterpart of the $B(E2)$ and half-lives except for $^{142}$Nd and $^{218}$U, whose half-lives show large discrepancies with the experimental values.
For $^{142}$Nd, the energy levels are reproduced well, but the $B(E2)$ transition value is very small (0.0003 W.u.). This could be the reason for the large magnitude of the calculated half-life. In the case of $^{218}$U, as there is no experimental information available for the electromagnetic transition of $8^+$, we can not say due to which part discrepancy is arising.

We have calculated $M1$ transition for the $8^+_2$ state of $^{218}$U, which is not measured experimentally. In comparison with the isomeric nature of $8^+_1$ state in $^{216}$Th, $\pi h_{9/2} f_{7/2}$ configuration has been proposed in \cite{218U} for the $^{218}$U. Following the behavior of $6^+$ states in the Sn-region and our shell-model results, we can propose the configuration of $8^+_2$ state for $^{218}$U to be based on $\pi h_{9/2}^2$ (or $\pi h_{9/2}^6 f_{7/2}^2 i_{13/2}^2$ ).  

\begin{figure*}
  \includegraphics[width=75mm]{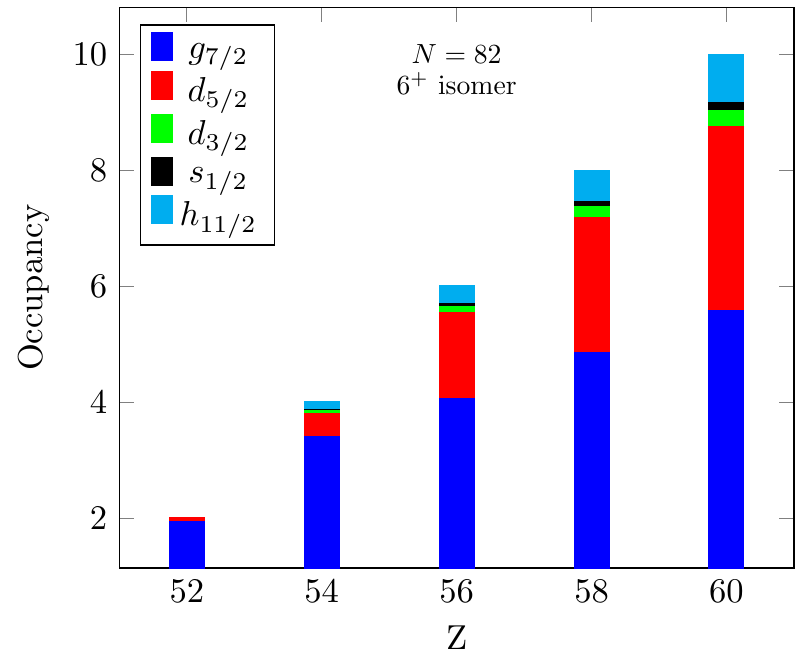}
  \includegraphics[width=75mm]{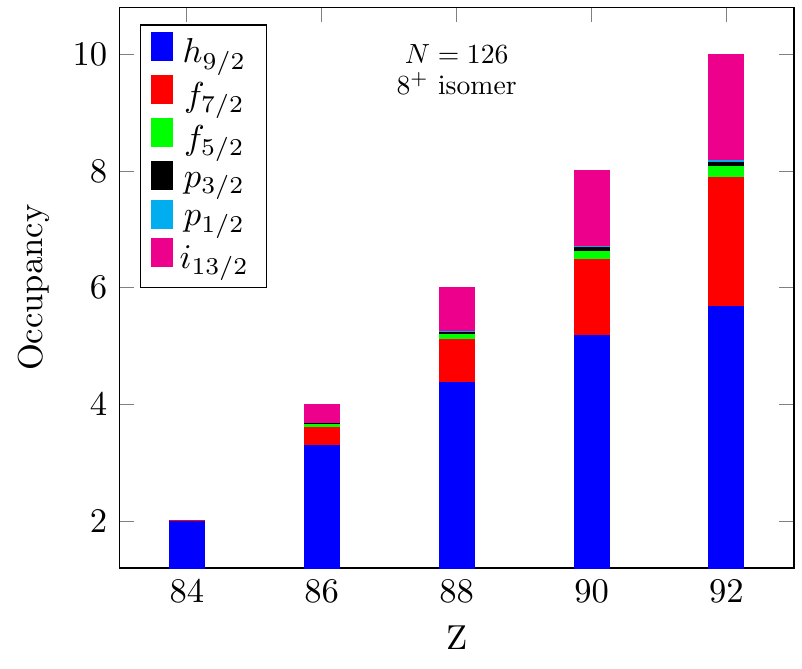}
  \label{fig:occupancy}
\caption{\label{occupancy}The occupancy for different orbitals corresponding to $6^+$ and $8^+$ isomers. }
\end{figure*}

\begin{figure*}
 \includegraphics[width=0.85\textwidth,height=1.7\textheight,keepaspectratio]{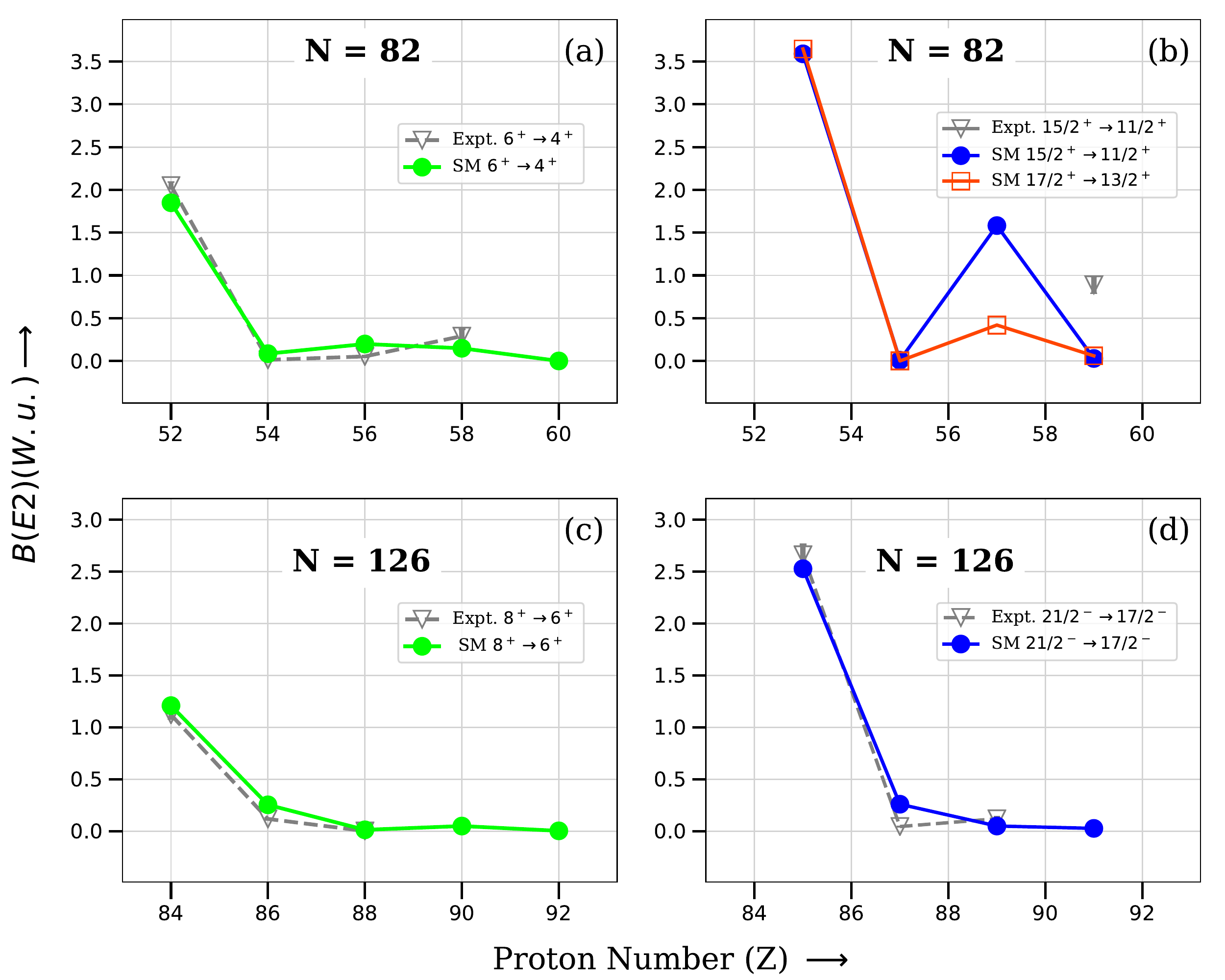}
 \caption{\label{BE2} Comparison of calculated and experimental \cite{nndc} $B(E2)$ values of different transitions for $N$ = 82 and $N$ = 126 isotones.}
\end{figure*}

In Fig. \ref{occupancy}, the occupancy of different orbitals for the $6^+$ isomers in $N$=82 and $8^+$ isomers in $N$=126 regions are shown. The behavior of the dominant part of the wave function is similar in both  the regions, i.e., the $\pi g_{7/2}$ orbital for $6^+$ isomers, while the $\pi h_{9/2}$ orbital for $8^+$ isomers. The contribution of $\pi h_{11/2}$ orbital in  the Sn-region and $\pi i_{13/2}$ orbital in the Pb-region is continuously increasing with the increase in $Z$ number. This shows the importance of these two high-$j$ orbitals even in the   formation of low-lying isomeric states, while the dominant orbitals  $\pi g_{7/2}$ and $\pi h_{9/2}$ are expected to be sufficient in the formation of low-lying isomeric states, giving a pure configuration. As we focus precisely on the contribution of different orbitals in the configuration, it appears that fragmentation of  the dominant wave function into different orbitals increases significantly with the increase in  the $Z$ number. This is reflected in the $B(E2)$ values of these two isomeric states discussed ahead. In Fig. \ref{BE2}, we can see a sharp decrease in the $B(E2)$ transition values for the $6^+$ and $8^+$ isomeric state, after $Z$ = 52 and 84, respectively. The $B(E2)$ values are higher at $Z$ = 52 and 84, as the configuration is pure, without any fragmentation into the other orbitals. There might be other factors affecting the behavior of the $B(E2)$ transition value. 


\begin{figure}
 \includegraphics[width=0.45\textwidth,height=1.7\textheight,keepaspectratio]{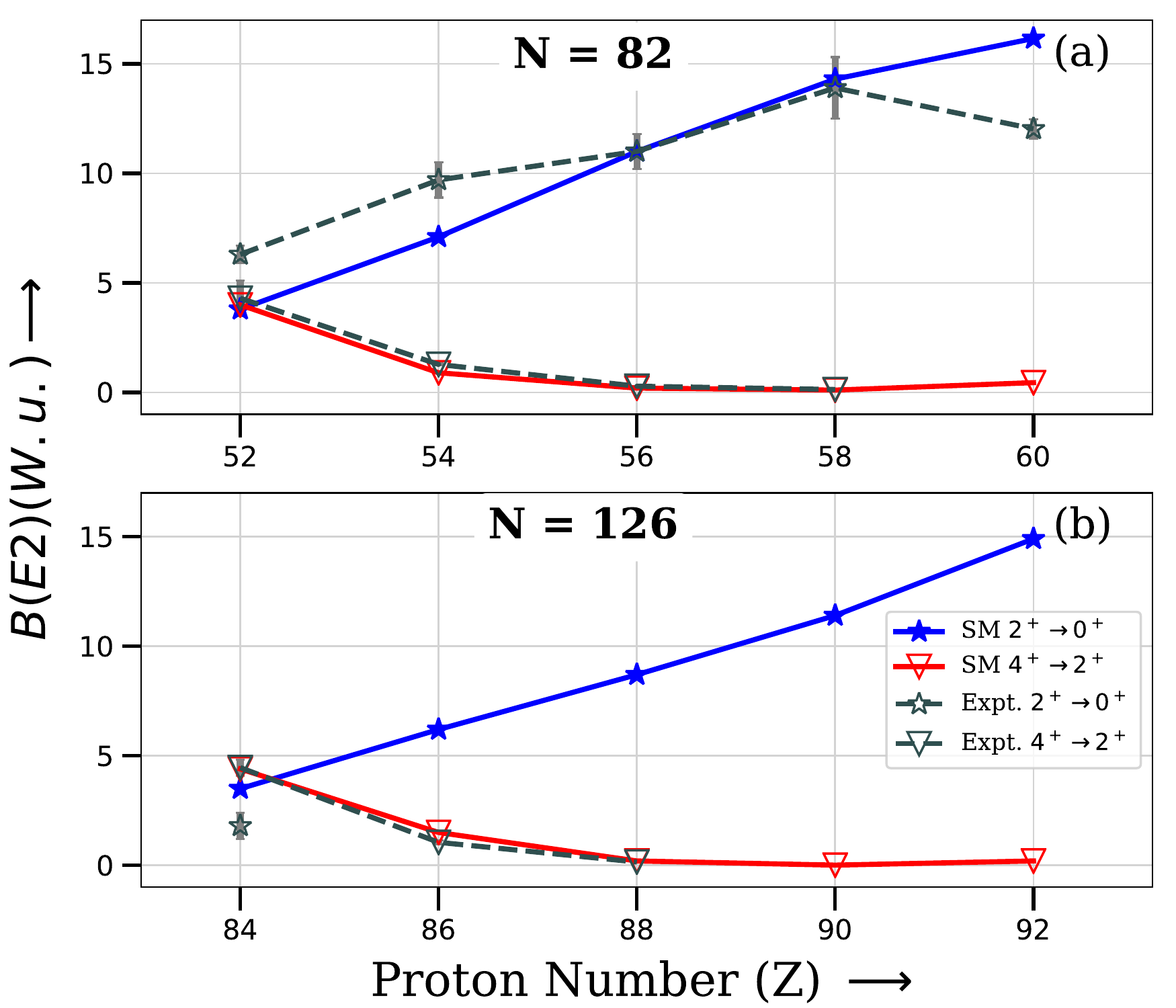}
 \includegraphics[width=0.46\textwidth,height=1.7\textheight,keepaspectratio]{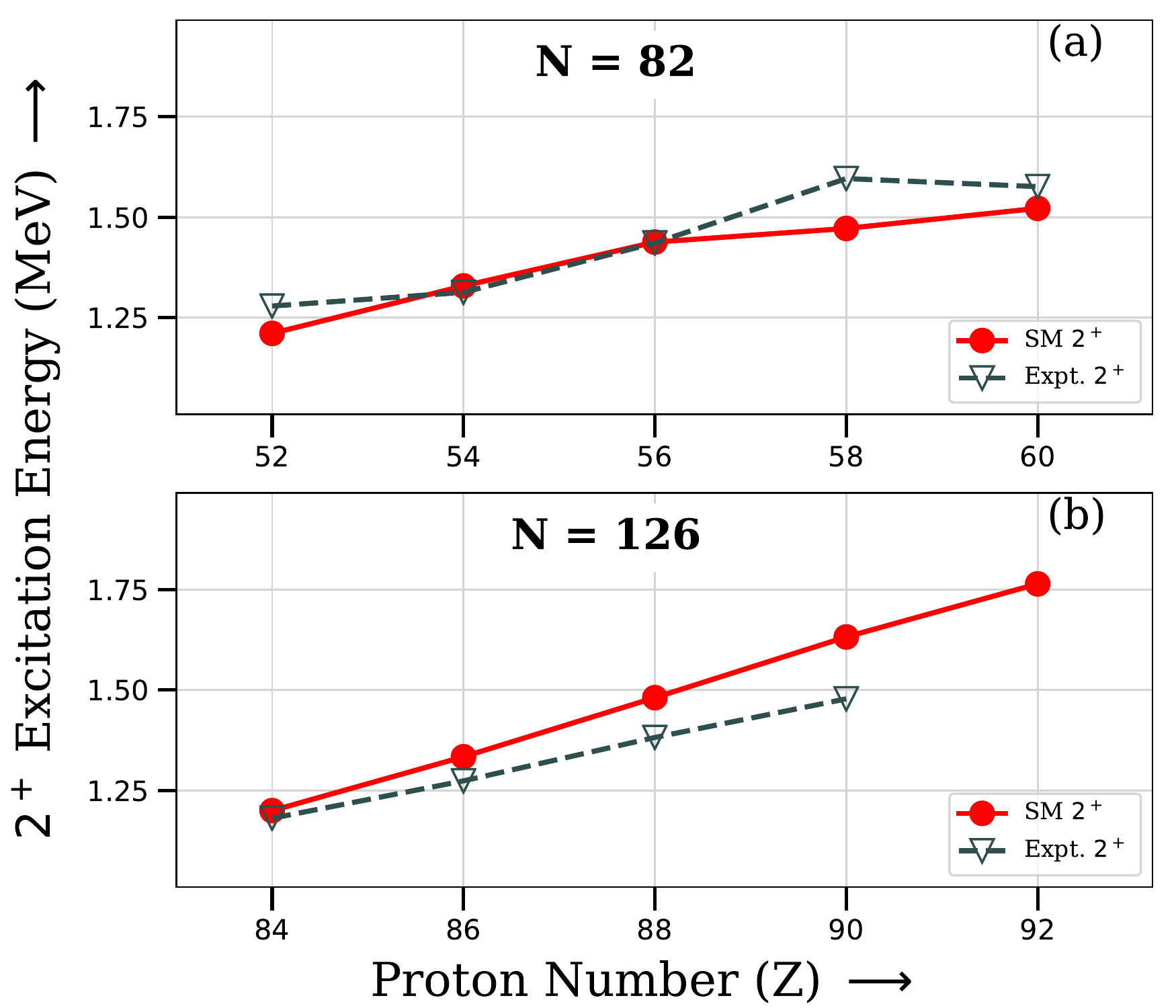}
 \caption{\label{2+BE2} Comparison of calculated and experimental \cite{nndc} $B(E2: 2^+ \rightarrow 0^+)$ and $B(E2: 4^+ \rightarrow 2^+)$ transition probabilities and first $2^+$ excitation energy for $N$ = 82 and $N$ = 126 isotones.}
\end{figure}

\subsection{$B(E2)$ and  $B(E3)$}

The reduced electric  transition probabilities provide much information. Such as, a small $B(E2)$ value indicates a single-particle structure, while a larger value provides an understanding of collective motion. It also tells about the accuracy of the shell-model wave functions.
Here, the $B(E2; J \rightarrow J - 2)$ transitions are calculated, interconnecting different even and odd parity states. The results for different transition rates, displayed in Fig. \ref{BE2} and the result of $B(E2: 2^+ \rightarrow 0^+)$ in Fig. \ref{2+BE2}. These values are calculated using the standard effective charges    $e_\pi = 1.5e$ and $e_\nu = 0.5e$.

\begin{table*}
\begin{center}
\begin{ruledtabular}
\caption{\label{table2}The calculated (SM) $B(E3)$ transition strengths in W.u. for Sn and Pb region isotones compared to the experimental data (Expt.) \cite{nndc}
using the effective charge $e_\pi = 1.5e$.}
\label{B(E3)}
\begin{tabular}{c|ccc|c|ccc|c|ccc}
$N$ = 82 & Transition & $B(E3)$ & $B(E3)$  & $N$ = 126 & Transition & $B(E3)$  & $B(E3)$ & $N$ = 126 & Transition & $B(E3)$  & $B(E3)$    \\
	     &            &   (SM)   &  (Expt.) &     Even  &            &  (SM)    & (Expt.) &    Odd   &           &  (SM)    & (Expt.) \\
	     \hline

{$^{134}$Te} & $9^-_1 \rightarrow 6^+_1$  &  1.43  &  3.80(14)  & {$^{210}$Po} & $11^-_1 \rightarrow 8^+_1$ & 0.34 & 3.71(10)  & $^{211}$At & ${29/2}^+_1 \rightarrow {23/2}^-_1$ & 7.71 & 21.5(19) \\
& $9^-_1 \rightarrow 6^+_2$ &  6.70 &  8.2(3)  &  & $11^-_1 \rightarrow 8^+_2$ & 7.62 & 19.7(11) &   & & & \\\hline

{$^{136}$Xe} & $9^-_1 \rightarrow 6^+_1$  &  1.52  &   & {$^{212}$Rn} & $11^-_1 \rightarrow 8^+_1$ & 0.22 & $1.8^{+6}_{-4}$  &  {$^{213}$Fr} & ${29/2}^+_1 \rightarrow {23/2}^-_1$ & 7.18 & 25.8(23)\\
 & $9^-_1 \rightarrow 6^+_2$ & 5.83 &   &   & $11^-_1 \rightarrow 8^+_2$ & 7.11 & $27^{+10}_{-7}$ &  & & & \\\hline

{$^{138}$Ba} & $9^-_1 \rightarrow 6^+_1$ & 5.72   &    & {$^{214}$Ra} & $11^-_1 \rightarrow 8^+_1$ & 0.19 & 3.09(9) &    {$^{215}$Ac} & ${29/2}^+_1 \rightarrow {23/2}^-_1$ & 6.52 & 27.4(9)\\
& $9^-_1 \rightarrow 6^+_2$ & 0.76 &    &  & $11^-_1 \rightarrow 8^+_2$ & 6.47 & 21.8(6) &  & & &\\\hline

{$^{140}$Ce} & $9^-_1 \rightarrow 6^+_1$ &  4.96  &    &  {$^{216}$Th} & $11^-_1 \rightarrow 8^+_1$ & 0.32 & 21(2) &   {$^{217}$Pa} & ${29/2}^+_1 \rightarrow {23/2}^-_1$ & 5.70 & \\
& $9^-_1 \rightarrow 6^+_2$ & 0.12 &   &   & $11^-_1 \rightarrow 8^+_2$ & 5.57  & 5.0(15) &  & & & \\\hline

{$^{142}$Nd} & $9^-_1 \rightarrow 6^+_1$  &  3.42  &    &  {$^{218}$U} & $11^-_1 \rightarrow 8^+_1$ & 4.45 &  & & & &  \\
& $9^-_1 \rightarrow 6^+_2$ & 0.04 &   & &   $11^-_1 \rightarrow 8^+_2$ & 0.55 &  & & &  &

		\end{tabular}
		\end{ruledtabular}
		
	\end{center}
\end{table*}

A similar general trend of decreasing $B(E2)$ value is observed in Fig. \ref{BE2} as the proton number increases for both the $N$ = 82 and 126 isotones. For the spins, $J \ge 2$ the decrement of $B(E2)$ values away from the shell-closure is an outcome of the seniority scheme. The $B(E2)$ transition that is even-tensor, one body transition between the states of the same seniority, will follow the trend determined by the matrix elements. Relating to this, the orbitals involved in the wave-functions of the initial and final states differ by one unit  of angular momentum and the formation of the involved states in transition rates for both regions follows the same seniority scheme. Surprisingly, the $B(E2)$ transition rates for the same seniority states in the two different closed-shell regions $N$ = 82 and 126 follow the same trend. The trend of $B(E2)$ transition rates suggests that seniority is a good quantum number for $N$ = 82 and 126 shell closure. This argument is only valid near the shell closure where valence nucleons are small because collectivity increases for a large number of valence particles and away from the shell closure.

In Fig. \ref{2+BE2}, we can observe increment of the $B(E2:2^+ \rightarrow 0^+)$ transition probabilities as the proton number increases. The large transition rates in $^{138}$Ba, $^{140}$Ce and $^{142}$Nd in the Sn-region and $^{216}$Th and $^{218}$U in the Pb-region indicate the collective nature of these nuclei. A weak $B(E2: 2^+ \rightarrow 0^+)$ transition rate is observed for the $^{134}$Te in the Sn-region similar to $^{210}$Po in the Pb-region. Overall, larger $B(E2)$ strengths are reached for larger $Z$ in the $N$ = 126 isotones.  Their strengths are similar as for the $N$ = 82 isotones in the $^{132}$Sn mass region.

From the results shown in Fig. \ref{2+BE2}, it is worth noticing that analogous increase in the first $2^+$ excitation energy is observed for $N$ = 82 and 126  regions as the proton number increases. For instance, 1.211 MeV in $^{134}$Te and 1.200 MeV in $^{210}$Po to 1.522 MeV in $^{142}$Nd and 1.764 MeV in $^{218}$U. The increase of $2^+$ and decrease of $4^+$ energy level and the larger $B(E2:2^+ \rightarrow 0^+)$ value than the $B(E2:4^+ \rightarrow 2^+)$ shows the spherical character of these nuclei in both the region. 
The increase in $B(E2: 2^+ \rightarrow 0^+)$ values is very rapid in both regions. In contrast, $2^+$ excitation energy shows a gradual increment. This behavior indicates the emergence of collectivity.

In Table \ref{table2} we have reported shell model results for $B(E3)$  transition strengths corresponding to the experimental data.
The octupole transitions in $N$ = 82 isotones resembles that for $N$ = 126 \cite{JP134Te, Bergstrom}. The $B(E3)$ transitions are retarded due to the spin-flip single particle transformation $\pi h_{11/2} \rightarrow \pi g_{7/2}$ and $\pi i_{13/2} \rightarrow \pi h_{9/2}$, in comparison with the transformation $\pi h_{11/2} \rightarrow \pi d_{5/2}$ and $\pi i_{13/2} \rightarrow \pi f_{7/2}$  which gives rise to fast $E3$ transition, in $N$ = 82 and $N$ = 126 isotones, respectively. Thus, the $B(E3: 9^-_1 \rightarrow 6^+_1)$ value increases with proton number (despite it decreases for $^{140}$Ce and $^{142}$Nd, because of the exchange of configuration), while the $B(E3: 9^-_1 \rightarrow 6^+_2)$ value decreases. Similar behaviour can be seen for $N$ = 126 isotones with weak $B(E3)$ transition. The $B(E3: 11^-_1 \rightarrow 8^+_1)$ value slows down with proton number and dominated by single proton spin-flip transitions (despite it increases for $^{216}$Th and $^{218}$U), while a strong $B(E3: 11^-_1 \rightarrow 8^+_2)$ value is reproduced.

The $11_1^-$ and $8^+_2$ are supposed to arise due to octupole coupling with $3^-$ core excitation. The $E3$ transition is expected to be very fast for the state generated from the coupling between $f_{7/2}$ orbital and the $3^-$ collective state. Thus,  strong $11_1^- \rightarrow 8^+_2$ transitions are observed for the  $11^-$ isomeric state,  which are due to single particle transformation $\pi i_{13/2} \rightarrow \pi f_{7/2}$. For $^{216}$Th, a fast $E3$ decay from $11^-$ state to $8^+_1$ state and a weak decay to the $8^+_2$ state  has  been observed experimentally. This is because the wave functions of the $8^+_1$ and $11^-$ should come from the contribution of $\pi f_{7/2}$, and mixing of $\pi h_{9/2}^7 f_{7/2}$ and $\pi h_{9/2}^8$ for the $8^+_2$ state. In contrast, in our shell-model calculation, the wave-function of the two $8^+$ states  is  reversed for $^{216}$Th, giving a weak decay to the $8^+_1$ state and fast decay for the $8^+_2$ state. In the case of $^{218}$U, our calculation replicates the observed behavior for the $^{216}$Th.

 In our shell-model calculation, the deficiency in the $B(E3)$ transitions is due to the need for larger model space, especially the particle-hole core-excitations across the 
shell gaps.  This could be examined in a different work by including a  larger model space in the shell model calculation \cite{PRL_Brown}.


\begin{figure*}
\begin{center}
 \includegraphics[width=0.7\textwidth,height=1.7\textheight,keepaspectratio]{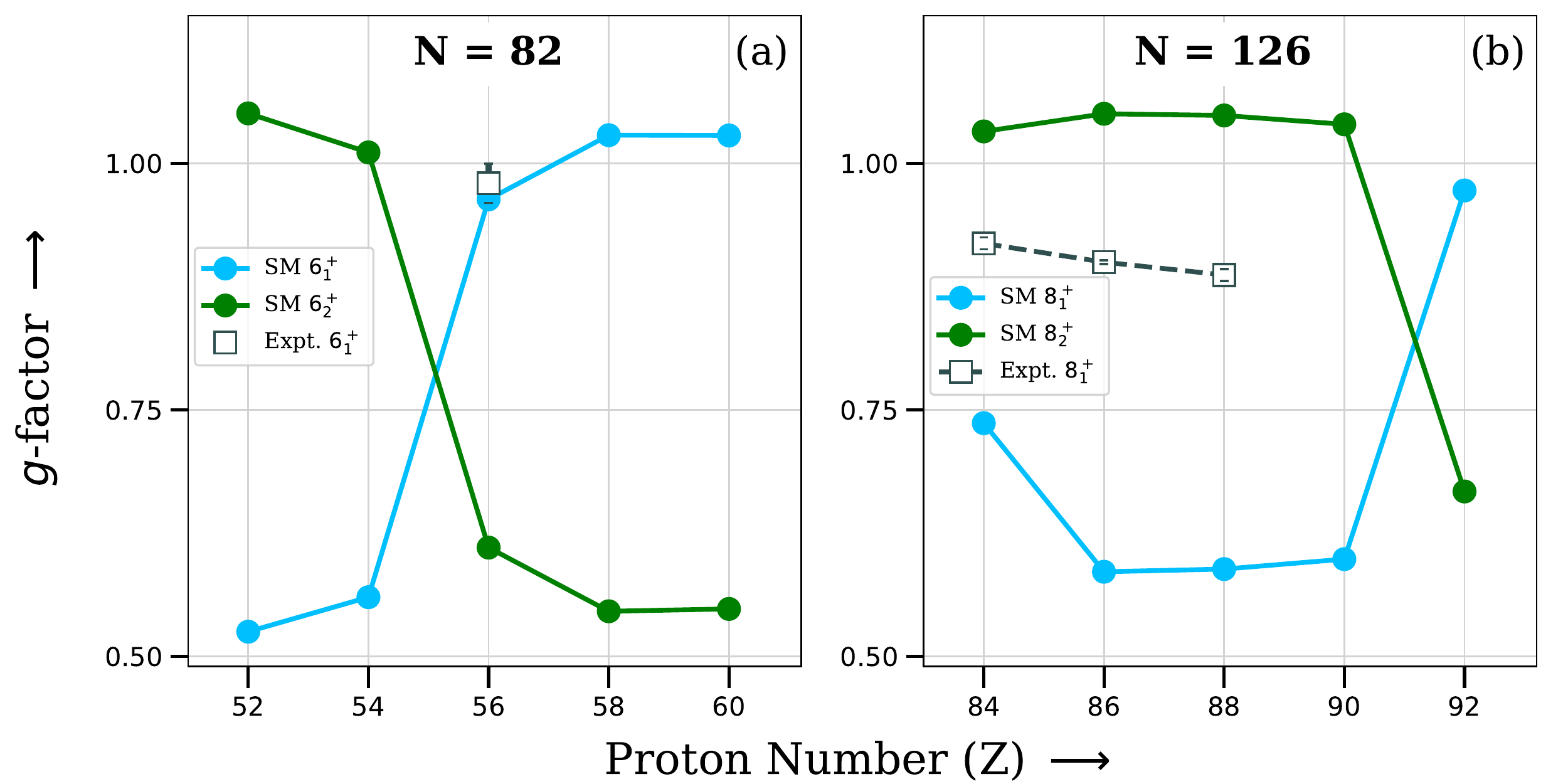}
 \caption{\label{gfactor} Trend of the $g$-factor for the two $6^+$ (in Sn region) and  $8^+$ (in Pb region) in comparison with the experimental data \cite{nndc}.}
 \end{center}
\end{figure*}

\begin{table}
\caption{ The calculated $g-$factor (in $n.m.$) and quadrupole moment (in $eb$) for isomeric states in $N=82$ and $N=126$ isotones compared to the experimental data \cite{nndc}.
 Shell-model calculations are performed with $g_{l}^\pi$=1 and $g_{s}^\pi$ = 5.585; $e_\pi = 1.5e$.}
 \label{g-factor}
\begin{center}
\begin{ruledtabular}
\begin{tabular}{c  c c  c   c c  c}
     A    & $J^\pi$   & $g-factor$  &  SM   & $Q (eb) $   &  SM \\
\hline 
$^{134}$Te & $6^+_1$  &         &   0.525  &             &  -0.373  \\
           & $9^-_1$  &         &   1.057  &             &  -0.673\\
$^{136}$Xe & $6^+_1$  &         &  0.560   &             &  -0.125 \\
           & $9^-_1$  &         &   1.055  &             &  -0.548 \\
$^{138}$Ba & $6^+_1$  & 0.98(2) &   0.964  &             &  -0.097 \\
           & $9^-_1$  &         &  1.056   &             &  -0.414 \\
$^{140}$Ce & $6^+_1$  &       &  1.029   &             &  0.120 \\
           & $9^-_1$  &         &  1.056   &             &  -0.315 \\
$^{142}$Nd & $6^+_1$  &       &  1.028   &             &  0.397 \\
           & $9^-_1$  &   +1.05 (13) &  1.056   &              &  -0.233 \\
\hline 
$^{135}$I & $17/2^+$  &       &   0.911   &   &  -0.569 \\
$^{137}$Cs & $17/2^+$  &       &   0.908   &   &  -0.267 \\
$^{139}$La & $17/2^+$ &        &   0.904  &   &  0.062 \\
$^{141}$Pr & $17/2^+$ &      &   0.896  &   &  0.411 \\

\hline 
$^{210}$Po & $8^+_1$  &   +0.919(6) &  0.737 & -0.552(20)  & -0.577  \\
           & $11^-_1$  &    &  1.024   &  -0.86(11)     & -0.967  \\
$^{212}$Rn & $8^+_1$  &  +0.894(2)   &  0.586 & -0.18(2)  & -0.308  \\
           & $11^-_1$  &    &  1.038   &             & -0.779  \\
$^{214}$Ra & $8^+_1$ &      0.885(4) & 0.589 &   & -0.060  \\
           & $11^-_1$  &  1.088(6)  &  1.038   &    &  -0.599 \\
$^{216}$Th & $8^+_1$ &  & 0.599 &   &  0.146 \\
           & $11^-_1$  &    &  1.037   &   &  -0.421 \\
$^{218}$U & $8^+_1$ &   & 0.973  &   &  0.035 \\
          & $11^-_1$  &    &  1.037   &  &  -0.248 \\

\hline
$^{211}$At & ${9/2}^-$  & +0.92 \cite{BDA} & 0.584  & -0.33(12)\cite{Cubiss} & -0.264 \\
           & ${21/2}^-$&  +0.910(8)  & 0.588  & 0.53(5)  &  -0.524  \\
$^{213}$Fr & ${9/2}^-$ &   +0.887(2)\cite{BDA}    & 0.585  &  -0.14(2) &  -0.121  \\
           & ${21/2}^-$  & 0.888(3)  &  0.5908  &     &  -0.178 \\
$^{215}$Ac & ${9/2}^-$  & +0.920(13) & 0.587  &  +0.04(10) &  0.093  \\
          & ${21/2}^-$   & 0.910(10) & 0.605 &   &  0.101 \\
$^{217}$Pa & ${9/2}^-$  &   	&   0.590 &   &  0.126 \\
        & ${21/2}^-$   &   & 0.795 &   &  0.194 \\
\end{tabular}
\end{ruledtabular}
\end{center}
\end{table}

\subsection{g-factor  and quadrupole moment}

In the Table \ref{g-factor}, we have reported shell model results for $g$- factor and quadrupole moments for the $6^+_1$, $9^-_1$ and $17/2^+$ isomers in $N$ = 82 isotones and $8^+_1$, $11^-_1$, ${21/2}^-$ isomers with ${9/2}^-$ ground state in $N$ = 126 isotones corresponding to the experimental data.

The $g$-factors are very sensitive to which orbital unpaired particles (or holes) are occupying.
Thus, the $g-$factor measurements explain the purity of a particular configuration and the contribution of the component in the wave function. The $g-$factors are particularly sensitive to the core polarization effects that couple the spin-orbit partners. 
The influence of $1p-1h$ excitation between spin-orbit (spin-flip) partners across a magic shell gap on the magnetic moment is called the first-order core polarization effect.
The influence of $2p-2h$ excitations is called second-order core polarization.
And the $g-$factors are not very sensitive to second-order core polarization effects (quadrupole particle-core coupling interactions). However, quadrupole moments are extremely sensitive to this. Therefore, the $g$-factor is found to be constant in a chain of isotopes or isotones.

According to the additivity rule, $g$-factors of the isomeric states with a rather pure $\pi (g_{7/2})^n$ or $\pi (h_{9/2})^n$ configuration should all be the same. Examples of the additivity of nuclear magnetic moments are given in Ref. \cite{Heyde}. The $9^-$ and $11^-$ isomers are from proton excitations out of the $Z$=50 and 82 core, respectively, so their $g-$factors are usually large and positive. These two isomers show almost the same $g-$factor throughout the isotonic chain. The $g-$factors of $9^-$ and $11^-$ isomers could be affected by particle-octupole vibration coupling admixtures. It needs to be investigated more from the theoretical perspective. In some isotones, due to first-order core polarization, Pauli-blocking can cause a breakdown of the additivity rule for $g$-factors. This can result in a changing $g$-factor as a function of $Z$. From our calculation we have seen that the configurations of the isomers are getting more mixed with increasing $Z$, in particular mixing with the $\pi (g_{7/2})^{n-1} d_{5/2}$ configuration in the Sn-region or $\pi (h_{9/2})^{n-1} f_{7/2}$ configuration in the Pb-region.

This mixing alone is not able to explain the breakdown of additivity. Other types of mixing are needed to be considered  in the wave function to describe the deviation from the additivity rule. This configuration mixing could be where the particle-core couple to octupole-vibrations and (or) first-order core polarization effects couple to valence particles. A detailed discussion can be found in \cite{Stuchbery}. The main impact of the octupole mixing into the wave function is that fewer protons will occupy the pure orbital. This results in a reduced Pauli blocking and thus better agreement with the experimental data.
The main configuration of the isomers is expected to be pure. The $g$-factor of these isomers should be the same for all of them (additivity theorem). 

The $g-$factors of $\pi g_{7/2}$ and $\pi h_{9/2}$  increase with the increasing proton number.
In the Sn-region, $g-$factor of one proton in the $g_{7/2}$ orbital is smaller than that of one proton in the $d_{5/2}$
orbital. Thus, when we take coupled $g$-factor with increasing $Z$ number, the $g$-factor of $6^+$ from the ($g_{7/2}$,$d_{5/2}$) configuration space will be larger than the ($g_{7/2}$,$g_{7/2}$). Therefore, the $g-$factor for the $6^+_1$ increases with the proton number while it decreases for the $6^+_2$.  In the Pb-region, the $g$-factor of $8^+$ from the ($h_{9/2}$,$f_{7/2}$) configuration space gives a larger $g$-factor than the ($h_{9/2}$,$h_{9/2}$). Similar to the Sn-region, the $g-$factor for the $8^+_1$ increases with the proton number while it decreases for the $8^+_2$. This shows the sensitivity of $g-$factors to the coupling of spin-flip partners (first-order core polarization) into the wave function. In Fig. \ref{gfactor} the evolution of the g-factor for two $6^+$ and $8^+$ states  has been shown with respect to the $Z$ number. From the figure, we can see that for the $8^+_1$ isomeric state, a deviation towards lower neutron numbers also appears.  
The more nucleons occupy the orbital, the higher the first-order core-polarization effect; thus deviation of $g$ - factor appears towards  a higher nucleon number. This parabolic curve might indicate that the correction in the $g$ - factor is related to the maximum number of particles (or holes).

Our calculated results of quadrupole moments, as shown in Table \ref{g-factor} are in a reasonable agreement with the available experimental data. Out of seven experimentally measured quadrupole moments, six quadrupole moments from the shell model are within the error bars and measured signs. In $^{211}$At, for  the $21/2^-$ state, the sign is not reproduced in our calculation. In the Sn-region, no experimental data is available for the quadrupole moments. Unlike the $g$-factor, quadrupole moments change in the isotonic chain with the increase in the $Z$ number.  
The deformation increases with the increase in the Z number above the shell-closure, and the quadrupole moment indicates the deformation. If we increase proton number beyond shell-closure, it is important to include the effect of second-order core polarization by including lower orbitals in shell-model space. 
The effect of second-order core polarization on the quadrupole moments can be seen in Table \ref{g-factor}. The deviation of the calculated quadrupole moment increases with the increase in the $Z$ number. 

\section{Conclusions}
\label{summary}

In the present work, we have performed a new type of investigation for the similarities between the $N=82$ ($52 \leq Z \leq 60$) and $N=126$ ($84 \leq Z \leq 92$) isotones in the framework of  the nuclear shell model with well known SN100PN and KHPE interactions, respectively. The isotones with $N$ = 82 and $N$ = 126 are semi-magic nuclei. 
From comparing the spectroscopy of Sn and Pb regions, similar patterns of nuclear structure are found in the framework of the shell-model.
The low-lying part of the spectrum should obey the seniority scheme, which is reflected in our present study. For these two regions, we have discussed spectroscopic and electromagnetic properties. Our study confirms several isomeric states in $N=82$ and $N=126$ isotones are due to the breaking of high-$j$ nucleon pairs and well described in terms of seniority quantum number. A similar evolution of excitation energies can be seen between the same seniority states in the Sn and Pb regions. We propose the wave function of $8^+_2$ state for $^{218}$U to be $\pi h_{9/2}^6 f_{7/2}^2 i_{13/2}^2$.

Using the correspondence between the Sn and Pb regions, the high-$j$ orbitals above the shell gaps show similar behavior, eventually deciding other spectroscopic properties. The decrease in $B(E2; J \rightarrow J - 2)$ transition value with proton number for the seniority $\nu$=2 and 3 multiplet is an outcome of  the seniority scheme and also due to the gradual decrease in the occupation of $\pi g_{7/2}$ and $\pi h_{9/2}$ orbitals in the Sn and Pb region, respectively. The rest of the occupation is scattered to the $d_{5/2}, h_{11/2}$ and  $f_{7/2}, i_{13/2}$ orbitals in Sn and Pb region, respectively.

 A strong resemblance can be seen in the growth of  $B(E2:2^+ \rightarrow 0^+)$ transition probabilities and in the variation of the first $2^+$ excitation energy. The $6^+_1$, $6^+_2$ and $9^-$ states in the Sn region and their counterpart $8^+_1$, $8^+_2$ and $11^-$ states in the Pb region  show several similarities  in the structural formation. Their evolution leads to a similar trend of $g-$factor and $B(E3)$ in both regions. Also, the structure of $9^-$ in the Sn region and $11^-$ states in the Pb region are related to the octupole coupling with two $6^+$ and $8^+$, respectively. Depending on the two different configurations, one among the two $6^+$ and $8^+$ states will be the octupole coupled core-excited state, which couples to $9^-$ and $11^-$ states in the Sn and Pb region, respectively. The $B(E3)$ value is crucial for a better understanding and explaining  the similarities between these two regions.
The retardation of the $B(E3)$ value in both regions is due to the spin-flip matrix element.
The breakdown in additivity for the $g-$factor could be explained with the first-order core polarization with mixing to the octupole coupling effects. These correlations can not be accounted in the present shell-model calculation.  Overall, shell model results agree with the available experimental data. We have also predicted electromagnetic properties where experimental data are not available. Our study will be beneficial  for future experimental study.

\section*{Acknowledgments}
B. Bhoy acknowledges financial support from MHRD, Government of India. P. C. Srivastava
acknowledges a research grant from SERB (India), CRG/2019/000556. 
We would like to thank Prof. Larry Zamick for several illuminating discussions.



\end{document}